\documentclass[prl,twocolumn,superscriptaddress,floatfix,10pt]{revtex4-1}
\usepackage{graphicx,bm}
\usepackage{amsmath}
\usepackage{amsfonts}
\usepackage{amssymb}
\usepackage{braket}
\usepackage{dsfont}
\usepackage{bm}
\usepackage{eucal}

\usepackage{color}

\usepackage{xcolor}
\usepackage[%
colorlinks=true,
urlcolor=blue,
linkcolor=blue,
citecolor=blue
]{hyperref}
\renewcommand{\vec}[1]{\boldsymbol{#1}}

\usepackage{tikz}
\usetikzlibrary{calc}
\usetikzlibrary{decorations.markings}
\usetikzlibrary{decorations.pathmorphing, arrows}
\usetikzlibrary{arrows.meta}

\usepackage{pdfpages} 
\makeatletter
\AtBeginDocument{\let\LS@rot\@undefined}
\makeatother

\newcommand{\eff}{{\text{eff}}}

\begin{document}
	\title{Hydrodynamic non-linear response of interacting integrable systems}
	\date{\today}
	\author{Michele Fava}
	\affiliation{Rudolf Peierls Centre for Theoretical Physics,  Clarendon Laboratory, Oxford OX1 3PU, UK}
	\author{Sounak Biswas}
	\affiliation{Rudolf Peierls Centre for Theoretical Physics,  Clarendon Laboratory, Oxford OX1 3PU, UK}
	\author{Sarang Gopalakrishnan}
	\affiliation{Department of Physics, The Pennsylvania State University, University Park, PA 16802, USA}
	\author{Romain Vasseur}
	\affiliation{Department of Physics, University of Massachusetts, Amherst, MA 01003, USA}
	\author{S. A. Parameswaran}
	\affiliation{Rudolf Peierls Centre for Theoretical Physics,  Clarendon Laboratory, Oxford OX1 3PU, UK}

	\begin{abstract}
		We develop a formalism for computing the non-linear response of interacting integrable systems. Our results  are asymptotically exact in the hydrodynamic limit where perturbing fields vary sufficiently slowly in space and time. 	
		We show that spatially resolved nonlinear response distinguishes interacting integrable systems from noninteracting ones, exemplifying this for the Lieb-Liniger gas. We give a prescription for computing finite-temperature Drude  weights of arbitrary order, which is in excellent agreement with numerical evaluation of the third-order response of the XXZ spin chain. 
		We identify intrinsically nonperturbative regimes of the nonlinear response of integrable systems.
	\end{abstract}

	\date{\today}
	
	\maketitle
	
	\textbf{
		Studying the response of a system to external fields yields information on its macroscopic order as well as its microscopic properties. While working to linear order in field strength often suffices to describe experiments, recent advances allow measurements to probe beyond the linear regime.  Understanding nonlinear response functions could significantly advance the characterization of exotic phases of matter, but little is known theoretically about their properties in interacting many-body systems. We introduce a framework for computing non-linear responses in a class of exactly solvable one-dimensional quantum systems. We show that non-linear response exhibits clear signatures of interaction effects, in contrast to linear response in similar settings.	
	}

	\section{Introduction}
	
		Most conventional experimental probes of many-body systems, from spectroscopy to transport, operate in the linear-response regime. Linear-response coefficients such as the a.c. conductivity and dynamical susceptibility have a natural theoretical interpretation in terms of the fluctuation-dissipation theorem~\cite{martin1968measurements}: the response to an external probe captures the intrinsic fluctuations of the system's degrees of freedom. Despite its many successes, linear response has its limitations as a probe of correlated quantum matter. For example, many different mechanisms --- of varying levels of interest --- give rise to incoherent spectral continua, and cannot be differentiated on the basis of linear-response data. Likewise, quantities like the conductivity probe some specific combination of the density and lifetimes of excitations; thus, e.g., the finite-frequency conductivity is qualitatively the same for a metal and an insulator. Recently, various experimental probes of \emph{nonlinear} response have been developed to circumvent these difficulties, ranging from quench experiments in ultracold atomic gases~\cite{bloch_review} to pump-probe spectroscopy~\cite{PhysRevLett.87.237401} and multidimensional coherent spectroscopy~\cite{mukamel1999principles,lynch2010first,lu2016nonlinear,hirori2011single,kuehn2011two,Jepsen01a,woerner2013ultrafast, Lu2017, mahmood2020observation, wan2019resolving, PhysRevLett.125.237601, nandkishore2020spectroscopic, nandkishore2021lifetimes, 2020arXiv201210603M, Jo_o_2019, PhysRevLett.124.117205, 2021arXiv210106081K, 2021arXiv210105019H, 2021arXiv210104136P} in condensed-matter settings. While the first of these methods is apt for probing far-from-equilibrium dynamics and the second radically reconstructs the state of the system, the third is milder, and probes higher-order and multiple-time correlations of the equilibrium system. 
		Such nonlinear probes are able to distinguish phases that have similar linear-response signatures: e.g., they can distinguish between excitation broadening due to disorder and that from decay~\cite{mahmood2020observation}. Despite a flurry of recent work~\cite{PhysRevLett.115.216806,wan2019resolving, PhysRevLett.125.237601,PhysRevB.99.045121, nandkishore2020spectroscopic,2020arXiv200107839L, nandkishore2021lifetimes, 2020arXiv201210603M, Jo_o_2019, PhysRevLett.124.117205, 2021arXiv210106081K, 2021arXiv210105019H, 2021arXiv210104136P}, the theoretical toolbox for addressing nonlinear response in generic interacting quantum many-body systems is primitive, with few exact results beyond free theories and those that reduce to ensembles of two-level systems. (Notable exceptions are Refs.~\cite{Doyon2019, 10.21468/SciPostPhys.8.1.007, 2020arXiv201206496P} which compute  specific time-ordered $n$-point correlation functions in integrable systems with the goal of characterizing ballistic transport.)
		
		\begin{figure}[bht]
		\centering
		\includegraphics[width=\linewidth]{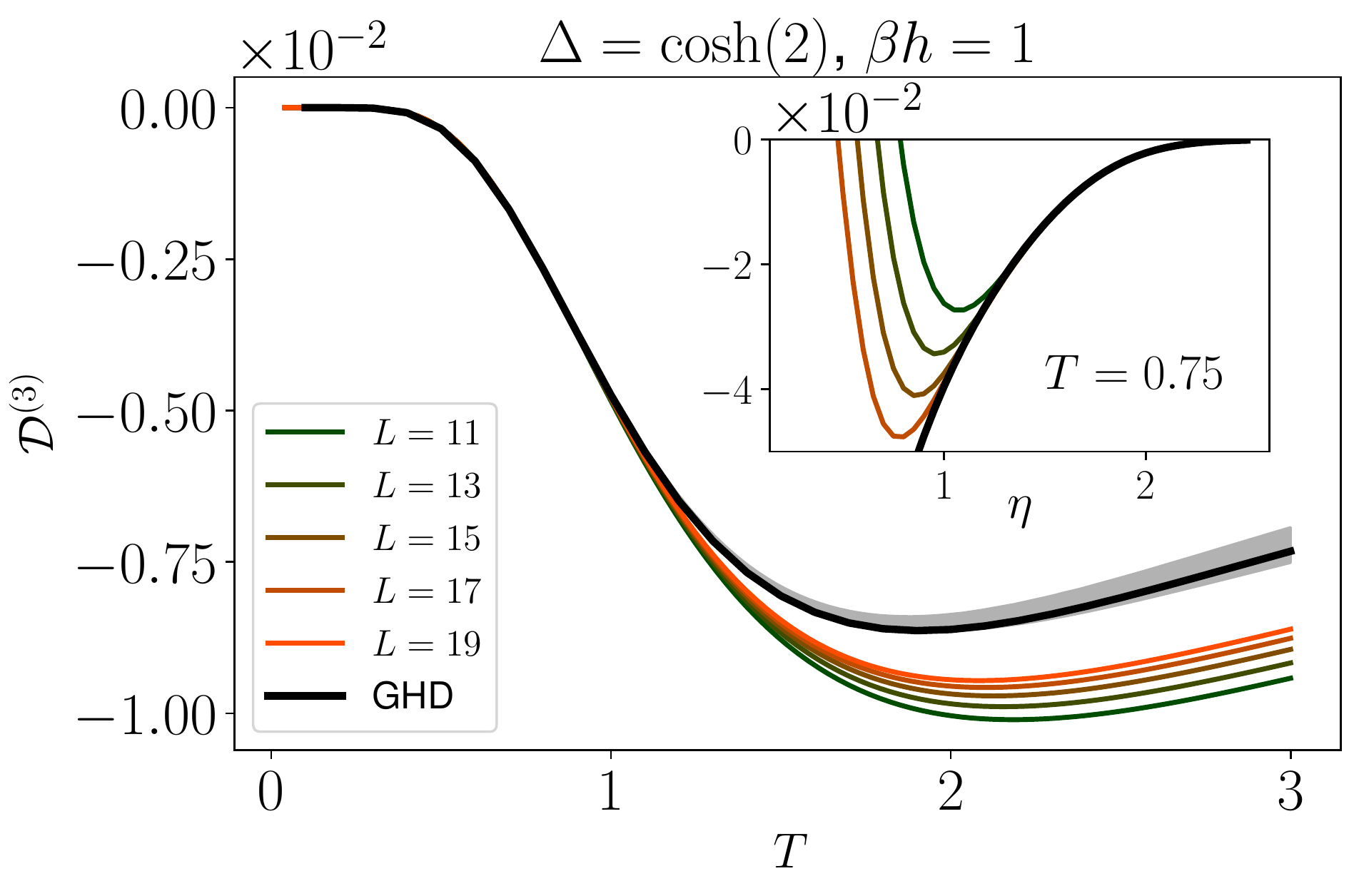}
		\caption{Third-order spin Drude weight $\mathcal{D}^{(3)}$ in the easy-axis regime of the XXZ spin chain with $\beta h=1$. Main figure: Comparison between GHD  and  ED results for fixed $\Delta$. The  lower (upper) boundaries of the shaded region correspond to extrapolations of finite-size ED results with a degree $1$ (degree $2$) polynomial in $1/L$.
			Inset: $\mathcal{D}^{(3)}$ as a function of $\eta=\cosh^{-1}\Delta$.}
		\label{fig:xxznumerics}
		\end{figure}
		
		Here, we develop and apply an asymptotically exact framework for computing the nonlinear response of interacting integrable systems, i.e. those solvable by
		the thermodynamic Bethe ansatz (TBA)~\cite{takahashi_book}. This framework is based on viewing integrability through the lens of generalized hydrodynamics (GHD)~\cite{Doyon, Fagotti, Doyon_notes} (see also~\cite{SD97,DS98} for a precursor of this approach, and {\it e.g.}~\cite{SciPostPhys.2.2.014,SciPostPhys.3.6.039,IN17, Bulchandani17,PhysRevB.96.115124,alba2017entanglement, PhysRevLett.120.045301,PhysRevLett.119.195301,PhysRevLett.120.164101, PhysRevLett.123.130602, DeNardis2018, Gopalakrishnan18, 2020arXiv200704861B, 10.21468/SciPostPhys.6.6.070, 2020arXiv200608577M, 2020arXiv201013738K,Pozsgay_algebraic,BPP20,PhysRevB.101.180302,durnin2020non,Fava20,SchemmerExpt,malvania2020generalized} for recent developments);
		our results are exact at the hydrodynamic Euler scale, i.e. for perturbations that vary slowly in space and time. {[The response to sharply localized potentials could contain oscillations in space and time that the GHD approach automatically averages out and hence cannot properly capture.]} {We remark that with these caveats GHD, and thus our method, is believed to be exact at any finite temperature and it can be applied to the computation of correlation functions of the density of any conserved charge in any integrable system.} 
		
		In the present work, we show that the nonlinear response of integrable systems contains information that is absent from (or subleading in) linear response: while the spectral functions of free and interacting integrable systems are qualitatively similar (with only subtle differences in the broadening around their ballistic light-cones~\cite{Gopalakrishnan18, DeNardis2018}), we find that spatially resolved nonlinear response reveals clear, qualitative distinctions between interacting and noninteracting integrable systems (as well as between chaotic and integrable systems). 
		We discuss the prospects for measuring these features in experiments on interacting many-particle systems using nonlinear spectroscopic probes.

		We also consider the generation of persistent currents after the application of an electric field, which is one of the hallmarks of integrability. At linear order in the field, the current is encoded in the linear Drude weight~\cite{SciPostPhys.3.6.039, PhysRevB.96.081118}. This can be readily generalized beyond the linear order, by defining higher-order Drude weights $\mathcal{D}^{(n)}$~\cite{PhysRevB.102.165137,Watanabe2020}. We show that our formalism yields a compact recursive formula for $\mathcal{D}^{(n)}$ at finite temperatures.
		We demonstrate the validity of our hydrodynamic approach by comparing its results with those of exact diagonalization studies of integrable spin chains; we find excellent agreement (Fig.~\ref{fig:xxznumerics}).
		We conclude by discussing the special case of the isotropic Heisenberg chain, which is known to host anomalous superdiffusive transport~\cite{Ljubotina_nature, Ilievski18, GV19, NMKI19, Ljubotina19, Vir20,NGIV20} characterized by propagation that is slower than  ballistic motion but faster than diffusion. We show that the emergence of supediffusion is accompanied by a breakdown of perturbation theory in the external field and hence an inherently non-perturbative nonlinear response.
		
	\section{Setup}
	
		We consider  general one-dimensional systems whose dynamics are governed by some integrable Hamiltonian $H_0$. 
		The dynamics under $H_0$ are treated within Euler-scale GHD~\cite{Doyon_notes}:
		we partition the system  into hydrodynamic cells each of mesoscopic size and linked to some spacetime point $(x,t)$, and assume that each cell is always instantaneously in some local generalized Gibbs ensemble (GGE)~\cite{Rigol07, ETH_review}, characterized by the vector of occupation factors of available quasiparticle states, $\vec{n}(x,t) = \{n_\theta(x,t)\}$; the ``rapidity'' $\theta$ is a convenient way of parameterizing the momentum. (We present  results for systems with a single quasiparticle species but the generalization to multiple species is immediate.) The density of quasiparticles of species $\theta$ can be expressed in terms of $\vec{n}$ as $\rho_\theta\equiv \rho^t_\theta n_\theta$, where $\rho^t_\theta$ is the available density of states for quasiparticles with those quantum numbers. (Note that, in an interacting system, $\rho^t_\theta$ is itself a nontrivial function of the local GGE.) Because of integrability, $\rho_\theta$ is separately conserved for each $\theta$; moreover, $n_\theta$ obeys a quasilinear advection equation, 
		\begin{equation}
		\label{eq:homogeneous-GHD}
		\partial_t n_\theta + v^{\mathrm{eff}}_\theta[\vec{n}] \partial_x n_\theta = 0, 
		\end{equation}
		where $v^\mathrm{eff}_\theta$ is an effective group velocity.
		In non-interacting systems, the effective velocity $v^\eff$ of a quasiparticle is just its group velocity. In an interacting integrable system, however, collisions are associated with a time delay in the quasiparticle trajectory, and thus renormalize the effective quasiparticle velocity. $v^\eff$ is therefore a nonlinear functional of $\vec{n}$.
		
		$H_0$ has an infinite set of conserved charges, $[H_0, \hat{Q}^j]=0$, whose expectation values in a GGE state are given by $\langle \hat{Q}^j\rangle = \int dx \langle \hat{q}_j\rangle = \int dx d\theta\, q^j_\theta \rho_\theta$, where $q^j_{\theta}$ is  the contribution to the $j^{\text{th}}$ charge density from quasiparticle $\theta$. The corresponding current density is $j_j = \int d\theta \rho_\theta q^j_\theta v^\mathrm{eff}_\theta$. GHD is highly nonlinear, even at the Euler scale, since the properties of each quasiparticle are strongly renormalized by its interactions with all the others; however, this nonlinearity can be
		addressed using TBA techniques.
		
		\begin{figure*}[t]
			\centering
			\begin{tikzpicture}
			\draw[step=0.5,gray,thin,dotted] (-0.50,0.5) grid (2.5,3.5);
			\draw[thick,->] (-0.5,0.5) -- (2,0.5) node[anchor=south west] {$x$};
			\draw[thick,->] (-0.5,0.5) -- (-0.5,3) node[anchor=south west] {$t$};
			
			\node (a) at (0.75,3.25) {(a)};
			\begin{scope}
			\node[shape=coordinate] (th1) at (0.4,0.6) {};
			\node[shape=coordinate] (preth1) at ($(th1)+(-0.05,-0.1)$) {};
			\node[shape=coordinate] (f1) at ($(th1)+(0.3,0.6)$) {};
			\node[shape=coordinate] (f2) at ($(f1) + (1, 1)$) {};
			\node[shape=coordinate] (m1) at ($(f2) + (0.2, 0.8)$) {};
			\node (O) at (1.75,3.25) {$\langle \hat{O}\rangle$};
			
			\draw[dashed] (preth1) -- ($(th1)+(0.1,0.2)$);
			\draw[thick] ($(th1)+(0.1,0.2)$) -- (f1) -- (f2) --(m1);
			
			\draw[color=red, thick ,decorate,decoration={snake,amplitude=.6mm}] (f1)-- ++(-1,0);
			\draw[color=red, thick ,decorate,decoration={snake,amplitude=.6mm}] (f2)-- ++(1,0);
			
			\draw[thick, color=gray] (0.5, 1) rectangle (1, 1.5);
			\draw[thick, color=gray] (1.5, 2) rectangle (2, 2.5);
			\draw[thick, color=gray] (1.5, 3) rectangle (2, 3.5);
			\end{scope}

			\begin{scope}[shift={(0.25,0)}]
			\node (b) at (4.75,3.25) {(b)};
			\draw[step=0.5,gray,thin,dotted] (2.99,0.5) grid (6,3.5);
			\node[shape=coordinate] (th1) at (3.6,0.7) {};
			\node[shape=coordinate] (preth1) at ($(th1)+(-0.1,-0.2)$) {};
			\node[shape=coordinate] (f1) at ($(th1)+(0.3,0.6)$) {};
			\node[shape=coordinate] (post-f1) at ($(f1) + (1.4, 1.4)$) {};
			\node[shape=coordinate] (post-post-f1) at ($(post-f1) + (0.6, 0.6)$) {};
			
			\draw[dashed] (preth1) -- ($(th1)+(0.05,0.1)$);
			\draw[thick] ($(th1)+(0.05,0.1)$) -- (f1) -- (post-f1);
			\draw[dashed] (post-f1) -- (post-post-f1);
			\draw[color=red, thick ,decorate,decoration={snake,amplitude=.6mm}] (f1)-- ++(-1,0);
			
			\node[shape=coordinate] (f2) at ($(th1) + (1.25, 1.45)$) {};
			\node[shape=coordinate] (th2) at ($(f2) + (0.15,-0.6)$) {};
			\node[shape=coordinate] (preth2) at ($(th2)+(0.15,-0.6)$) {};
			\node[shape=coordinate] (m2) at ($(3.9,3)$) {};
			\node (O) at (3.75,3.25) {$\langle \hat{O}\rangle$};

			\draw[dashed, color=blue] (preth2) -- (th2);
			\draw[thick, color=blue] (th2) -- (f2) -- (m2);
			\draw[color=red, thick ,decorate,decoration={snake,amplitude=.6mm}] (f2)-- ++(1,0);
			
			\draw[thick, color=gray] (3.5, 1) rectangle (4, 1.5);
			\draw[thick, color=gray] (4.5, 2) rectangle (5, 2.5);
			\draw[thick, color=gray] (3.5, 3) rectangle (4,3.5);
			\end{scope}

			\begin{scope}[shift={(0,0)}]
			\node (c) at (7.75,3.25) {(c)};
			\draw[step=0.5,gray,thin,dotted] (6.99,0.5) grid (10.5,3.5);
			\node[shape=coordinate] (th1) at (7.5,0.6) {};
			\node[shape=coordinate] (f1) at ($(th1)+(0.2,0.6)$) {};
			\node[shape=coordinate] (actual-th1) at ($(f1)+(-0.05, -0.3)$) {};
			\node[shape=coordinate] (preth1) at ($(actual-th1)+(-0.066,-0.4)$) {};
			\node[shape=coordinate] (post-f1) at ($(f1) + (1, 1.8)$) {};
			
			\draw[dashed] (preth1) -- (actual-th1);
			\draw[thick] (actual-th1) -- (f1) -- (post-f1);
			\draw[color=red, thick ,decorate,decoration={snake,amplitude=.6mm}] (f1)-- ++(-1,0);
			\draw[thick, color=gray] (7.5, 1) rectangle (8, 1.5);
			
			\node (O) at (8.75,3.25) {$\langle \hat{O}\rangle$};
			\draw[thick, color=gray] (8.5, 3) rectangle (9,3.5);
			
			\node[shape=coordinate] (f2) at ($(th1) + (2.2, 1.2)$) {};
			\node[shape=coordinate] (th2) at ($(f2) + (0.15,-0.6)$) {};
			\node[shape=coordinate] (preth2) at ($(th2)+(0.15,-0.6)$) {};
			\node[shape=coordinate] (m2) at ($(post-f1)+(0.2,0)$) {};
			
			\draw[dashed, color=blue] (preth2) -- (th2);
			\draw[thick, color=blue] (th2) -- (f2) -- (m2);
			\draw[color=red, thick ,decorate,decoration={snake,amplitude=.6mm}] (f2)-- ++(1,0);
			\draw[thick, color=gray] (9.5, 1.5) rectangle (10,2);
			\end{scope}

			\begin{scope}[shift={(0.25,0)}]
			\node (d) at (12.75,3.25) {(d)};
			\draw[step=0.5,gray,thin,dotted] (10.99,0.5) grid (14.5,3.5);
			
			\node[shape=coordinate] (th1) at (11.5,0.6) {};
			\node[shape=coordinate] (preth1) at ($(th1)+(-0.05,-0.1)$) {};
			\node[shape=coordinate] (f1) at ($(th1)+(0.3,0.6)$) {};
			
			\node[shape=coordinate] (c1) at ($(f1)+(0.8,0.8)$) {};
			\node[shape=coordinate] (c2) at ($(f1)+(1.2,1.1)$) {};
			
			\node[shape=coordinate] (post-f1) at ($(f1) + (1.9, 1.8)$) {};
			\node[shape=coordinate] (post-post-f1) at ($(post-f1) + (0.4, 0.4)$) {};
			
			\draw[dashed] (preth1) -- ($(th1)+(0.1,0.2)$);
			\draw[thick] ($(th1)+(0.1,0.2)$) -- (f1) -- (c1);
			\draw[thick, dotted] (c1) --(c2);
			\draw[thick] (c2) -- (post-f1);
			\draw[dashed] (post-f1) -- (post-post-f1);
			\draw[color=red, thick ,decorate,decoration={snake,amplitude=.6mm}] (f1)-- ++(-1,0);
			\draw[thick, color=gray] (11.5, 1) rectangle (12, 1.5);

			\node (O) at (11.75,3.25) {$\langle \hat{O}\rangle$};
			\draw[thick, color=gray] (11.5, 3) rectangle (12,3.5);
			
			\node[shape=coordinate] (f2) at ($(th1) + (2.2, 1.1)$) {};
			\node[shape=coordinate] (th2) at ($(f2) + (0.15,-0.6)$) {};
			\node[shape=coordinate] (preth2) at ($(th2)+(0.15,-0.6)$) {};
			\node[shape=coordinate] (c1) at ($(f2)+(-0.7,0.35)$) {};
			
			\node[shape=coordinate] (m2) at ($(O)+(0,-0.25)$) {};
			\node[shape=coordinate] (c2) at ($(m2)+(1,-0.5)$) {};

			\draw[dashed, color=blue] (preth2) -- (th2);
			\draw[thick, color=blue] (th2) -- (f2) -- (c1);
			\draw[thick, dotted, color=blue] (c1) -- (c2);
			\draw[thick, color=blue] (c2) -- (m2);
			\draw[color=red, thick ,decorate,decoration={snake,amplitude=.6mm}] (f2)-- ++(1,0);
			
			\draw[thick, color=gray] (12.5, 2) rectangle (13, 2.5);
			\draw[thick, color=gray] (13.5, 1.5) rectangle (14, 2);

			\end{scope}

			\end{tikzpicture}
			\caption{
				Four distinct physical processes contributing to the second-order response $\chi^{(2)}$.
				(a) A thermal quasiparticle (QP; black line) is accelerated twice by the electric field (red wavy line), and modifies the expectation value $\langle \hat{O} \rangle$ in the final space-time cell.
				(b) First a thermal QP is accelerated; a second thermal QP (blue line) is later accelerated when the first is in its space-time cell, thus modifying the effective acceleration perceived by the second;  both QPs proceed ballistically and the second modifies $\langle \hat{O} \rangle$.
				(c) Two thermal QPs are independently accelerated by two pulses of the electric field; after travelling to the same space-time cell, and together they modify $\langle \hat{O} \rangle$.
				(d) As in (c), two thermal QPs are independently accelerated but one scatters off the other before influencing $\langle\hat{O}\rangle$. Only  (a) is relevant to free systems but all four processes contribute in  interacting integrable systems.
			}
			\label{fig:diagrams-chi2}
		\end{figure*}
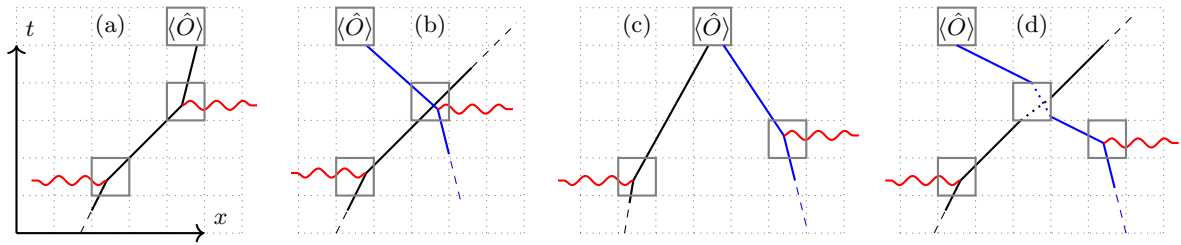
		
		We now discuss how external forces can be incorporated into GHD~\cite{SciPostPhys.2.2.014, PhysRevLett.123.130602}.	
		For concreteness we specialize to the case where the coupling is to a global $U(1)$  charge $\hat{q} = \hat{q}_0$, which remains conserved even in the presence of inhomogeneous fields. Thus, the perturbed Hamiltonian is $
		{H}(t) = \hat{H}_0 + \int dx\,V(x,t) \hat{q}_0(x)$.
		Assuming $V$ varies  slowly in space and time, the Euler-scale time evolution of the system is described by~\cite{SciPostPhys.2.2.014, PhysRevLett.120.164101, PhysRevLett.123.130602, 2020arXiv200704861B, 10.21468/SciPostPhys.6.6.070, 2020arXiv200608577M, 2020arXiv201013738K}
		\begin{equation}
		\label{eq:inhomogeneus-GHD}
		\partial_t n_\theta + v^\eff_\theta \partial_x n_\theta
		+ E a_\theta^\eff \partial_\theta n_\theta=0,
		\end{equation}
		where 
		$a_\theta^\eff [\vec n]$
		is the effective acceleration of the quasiparticles, and
		the sole dependence on the potential is via the electric field $E(x,t) \equiv -\partial_x V(x,t)$. As is the case for $v^\eff$, $a^\eff$ is also renormalized by scattering processes and is hence a nonlinear functional of $\vec{n}$. 
		
		Finally, we note that \eqref{eq:inhomogeneus-GHD} is strictly valid only at the Euler scale, i.e., for response at asymptotically large $x$ and $t$, but with a fixed ratio $x/t$. Euler-scale response is a hallmark of integrable dynamics: chaotic systems without Galilean invariance have exponentially suppressed response at the Euler scale, since densities spread diffusively rather than ballistically. Interacting integrable systems also have diffusive corrections to ballistic quasiparticle spreading~\cite{DeNardis2018, Gopalakrishnan18, DeNardis_SciPost}, but these corrections are also suppressed at the Euler scale.
		
	\section{Nonlinear-response}
	
		Response is concerned with computing the value of some local observable $\hat O$---taken here to be a charge density  $\hat{q}_j$ or current density $\hat{j}_j$---following the application of electric fields $E(x,t)$.
		Since~\eqref{eq:inhomogeneus-GHD} is asymptotically {exact} at the Euler scale  to all orders in $V_j$, it is sufficient to work perturbatively in $V_j$ to compute the response (we comment on exceptions below).
		Formally, the connected order-$N$ response is
		\begin{equation}
		\label{eq:chiN}
		\chi^{(N)}_{\hat{O}} \left(\{x_n, t_n \}\right) = \left.\prod_{n=0}^{N-1}
		\frac{\delta}{\delta E(x_n, t_n)} \langle \hat{O}(x_N,t_N)\rangle\right|_{E\to 0} 
		\end{equation}
		with $t_0<t_1<\cdots<t_N$.
		The expectation value  is taken with respect to the nonuniform state at time $t_N$  generated by perturbing the  initial uniform GGE state with external fields  at times $t_1, \ldots t_{N-1}$. 
		Our strategy  is to express the expectation value in  \eqref{eq:chiN} in terms of quasiparticle  occupations,
		perform all the functional derivatives, and then  set $E=0$, yielding an  expression that we  evaluate in the {\it uniform} GGE.
		
		An expectation value $\langle O(x, t) \rangle$ is a nonlinear functional of the local state $\vec{n}(x, t)$. It can be affected by perturbations at other spacetime points only through the advection of those perturbations to $(x,t)$, which is captured by the propagator $D_{\theta\theta'}(z,z')  = \frac{\delta n_\theta(z)}{\delta n_{\theta'}(z')}$,	where we have defined $z \equiv (x, t)$. 	One can express this dependence in the following compact form, suggestive of a chain rule~\cite{10.21468/SciPostPhys.5.5.054}:
		\begin{equation}
		\label{eq:factorization}
		\frac{\delta \langle \hat{O}(z_1) \rangle}{\delta E(z_0)} = \!\int\!\! d\theta d\alpha \frac{\delta  \langle \hat{O} (z_1) \rangle}{\delta n_\alpha(z_1)}  \frac{\delta n_\alpha(z_1)}{\delta n_\theta(z_0)} \frac{\delta n_\theta(z_0)}{\delta E(z_0) },
		\end{equation}
		\eqref{eq:factorization} simply says that expectation values at spacetime point $z_0$ depend on  fields at $z_1\neq z_0$
		purely via the process by which the fields perturb the quasiparticle distribution at $z_0$ and this perturbation is advected over to $z_1$. 
		
		We are interested in generalizing~\eqref{eq:factorization} to the case of higher-order functional derivatives. 
		To organize these more complicated expressions we have developed a diagrammatic framework~\cite{SM}, which relies on the observation that any functional derivative can be composed of the following types of elementary object. First, there are propagators, defined above, connecting perturbations at different spacetime points; in a uniform GGE, the propagators take the simple form $D_{\theta\theta'}(x_0, t_0; x_1, t_1) = \delta_{\theta\theta'} \delta\left[(x_0 - x_1) - v^\mathrm{eff}_\theta (t_0 - t_1)\right]$~\cite{Doyon_correlations}. Second, there are functional derivatives of observables at a point with respect to the quasiparticle distribution at the same point, which can be evaluated using TBA techniques~\cite{Doyon_notes}. We call these ``measurement vertices.'' Third, there are derivatives of the quasiparticle distribution at a spacetime point with respect to fields at the same point. To find these we invert \eqref{eq:inhomogeneus-GHD} using Green's function techniques, and thereby find  $\frac{\delta n_\theta (z)}{\delta E(z)} = -a^{\text{eff}}_\theta[\vec{n}] \partial_\theta {n}_\theta$~\cite{SM}. We term these objects  ``field vertices.'' These three types of objects appear in~\eqref{eq:factorization}. Finally,
		response functions at order  $N>1$ will also involve expressions of the form $\Gamma^{(p)}=\frac{\delta^p n_\theta(z_0)}{\delta n_{\theta_1}(z_1)\ldots \delta n_{\theta_p}(z_p)}$. These capture the modification of the spacetime propagator by scattering events, and can be computed by repeatedly differentiating~\eqref{eq:homogeneous-GHD} 
		with respect to $n$,  which yields a recursive formula, that allows us to express $\Gamma^{(p)}$  in terms of $\partial_x\Gamma^{(1)}$ and functional derivatives of the $v^{\text{eff}}[\vec{n}]$ with respect to quasiparticle occupations~\cite{SM}. We refer to these objects as ``scattering vertices.''
		All other types of object can be expressed in terms of  these: e.g., functional derivatives of the form $\frac{\delta^k \langle\hat{O}\rangle}{\delta n_{\theta_1}(z_1)\ldots \delta n_{\theta_k}(z_k)}$, can be rewritten in terms of measurement or scattering vertices and propagators, which advect all  occupation factors to the point where the functional derivative is taken. We may verify that for $N=1$ this procedure yields the standard expressions for  linear response. Higher-order response functions can then be computed recursively from \eqref{eq:chiN}.

		Although the formal expressions rapidly become unwieldy with increasing $N$, they have a transparent physical interpretation, as we now exemplify for $N=2$. The external field  can affect the system via two distinct physical processes, each corresponding to  a distinct field vertex (represented by a box with a wavy line in Fig.~\ref{fig:diagrams-chi2}): it can accelerate a thermal quasiparticle from rest within a spacetime cell (the first field vertex in Fig.~\ref{fig:diagrams-chi2}a), or else accelerate a quasiparticle  previously acted upon by the field at an earlier time (the second field vertex in Fig.~\ref{fig:diagrams-chi2}a). In a non-interacting integrable system different quasiparticles are independent of each other, thus all connected nonlinear response functions result solely when a single quasiparticle is repeatedly accelerated by the field, and then measured, as in Fig.~\ref{fig:diagrams-chi2}a.
		However, in interacting integrable systems, quasiparticles influence each other via  scattering processes.
		Consequently, the ability of the field to excite a quasiparticle in a given spacetime cell $z$ is also sensitive to the presence of quasiparticles excited by the field in all spacetime cells in the past light-cone of $z$ under the advective dynamics of GHD, leading to additional connected contributions (as in Fig.~\ref{fig:diagrams-chi2}b).  Quasiparticles excited by the field acting at distinct spacetime cells can also propagate to a single cell where they jointly modify the measured observable (Fig.~\ref{fig:diagrams-chi2}c). Interactions thus lead to an infinite hierarchy of  field and measurement vertices, that are sensitive to the presence of an increasing number of previously-excited quasiparticles in the spacetime cells where quasiparticles are accelerated or measured.  Finally, the  nonlinear response also receives contributions from scattering vertices, again of arbitrary order, due to the phase shift experienced by the measured quasiparticle as it propagates between the acceleration and measurement cells in the presence of  other excited quasiparticles in the system (Fig.~\ref{fig:diagrams-chi2}d). The  $N^{\text{th}}$ order response in an interacting integrable system involves $N$ field vertices and a single measurement vertex, linked by advection propagators $D_{\theta\theta'}(z,z')$ and scattering vertices, and can be organized  using spacetime diagrams~\cite{SM}. Crucially, at fixed $N$, only vertices below some finite order can contribute: for instance Fig.~\ref{fig:diagrams-chi2} contains all processes contributing to $\chi^{(2)}$.
		
		We caution the reader that in Fig.~\ref{fig:diagrams-chi2} the effects of fields and collisions are exaggerated for clarity. In fact, the trajectory shift due to scattering processes as in Fig.~\ref{fig:diagrams-chi2}d is infinitesimal, and similarly a perturbing external field only imparts an infinitesimal acceleration to each quasiparticle. Thus there are kinematic restrictions on allowed processes that the figure does not capture. For instance, the process in Fig.~\ref{fig:diagrams-chi2}a is possible only if the three points --- the two where the field act and the one at which the measurement occurs  --- lie on the same ray { $x=x_0 + v_\lambda t$ for some initial position $x_0$ and some rapidity $\lambda$}. This aspect will be crucial to our discussion in the next section.
		
	\section{Measuring Interactions in the Lieb-Liniger Gas}
	
		As an example of this approach, we apply it to the Lieb-Liniger model of 1D bosons with contact interactions,
		\begin{equation}
		\hat{H}_0 = \frac{1}{2} \sum_j \hat{p}_j^2 + c \sum_{i\neq j} \delta(\hat{x}_i - \hat{x}_j),
		\end{equation}
		where  $\hat{x}_j$ and $\hat{p}_j$ are the position and momentum of particle $j$.
		The bare group velocity $v$ of a particle is equal to its momentum $p$. The effective velocity $v^\eff$ can be obtained from $v$ as the solution to an integral equation, whose explicit form is provided in the Methods section. An additional fact,  peculiar to the Lieb-Liniger gas,  is that the effective acceleration $a^\eff$ is not renormalized by interactions; with our choice of conventions, $a^\eff=1$.
		For $c\to 0$, $\hat{H}_0$ is a free Bose gas, while for $c\to\infty$ it can be described as a theory of free fermions.
		This can be  recognized, for example, by studying $v^\eff$, which in both limits tends to the bare group velocity $v$.
		Consequently {\it linear} response  in these two limits approximates that of free bosons or  fermions respectively, with only quantitative corrections from interactions. This hinders a precise measurement of  $c$ based only on linear response.
		We now demonstrate that a spatially-resolved measurement of $\chi^{(2)}$ ---or higher-order responses--- carries direct information about the interactions.
		For concreteness, we consider a specific charge response of the form  $\tilde{\chi}^{(2)}(x,t,\tau) \equiv \chi^{(2)}_{\hat{q}_0}(0, 0; x, \tau; 0, \tau+t)$  where the first perturbation and the measurement  coincide spatially, and the system is perturbed at an intermediate time at position $x$.
		In the free boson or free fermion limits, we know from the discussion in the previous section that the only process contributing to $\chi^{(2)}$ is  one where a single quasiparticle is repeatedly accelerated by the subsequent field applications (Fig.~\ref{fig:diagrams-chi2}a). Furthermore, as previously noted, this process can take place only if all the points in which the perturbation is applied and the measurement point lie on the same ray. Thus, in the two non-interacting limits $\tilde{\chi}^{(2)}(x,t,\tau)$
		will vanish everywhere except at $x=0$.
		Conversely, if $c=O(1)$, quasiparticles are strongly interacting, and each influences the dynamics of the others. For example,  processes such as those in Fig.~\ref{fig:diagrams-chi2}d will be non-zero since $v^\eff$ of a quasiparticle with momentum $p$ will depend on all the quasiparticles in the same region.~\cite{SM} We thus expect that $\chi^{(2)}$ is generically finite and non-zero for arbitrary perturbation and measurement points.
		
		To summarize: if we focus on the region away from $x=0$, i.e. chosen to exclude the case where  all points lie along the same ray, we  expect $\tilde{\chi}^{(2)}(x,t,\tau)$ to be directly sensitive to the interactions, and hence generically will have a nonzero value away from the free limits  $c\to0$ or $c\to\infty$. An immediate corollary is that in these limits,  $\tilde{\chi}^{(2)}(x\neq 0,t,\tau)$  should respectively vanish as $O(c)$ or $O(1/c)$. 
		This should be contrasted with linear response measurements where $\chi^{(2)}=O(1)$ in all these cases and the effect of interactions is to determine sub-leading corrections.
		
\begin{figure}[hbt]
			\centering
			\includegraphics[width=0.98\linewidth]{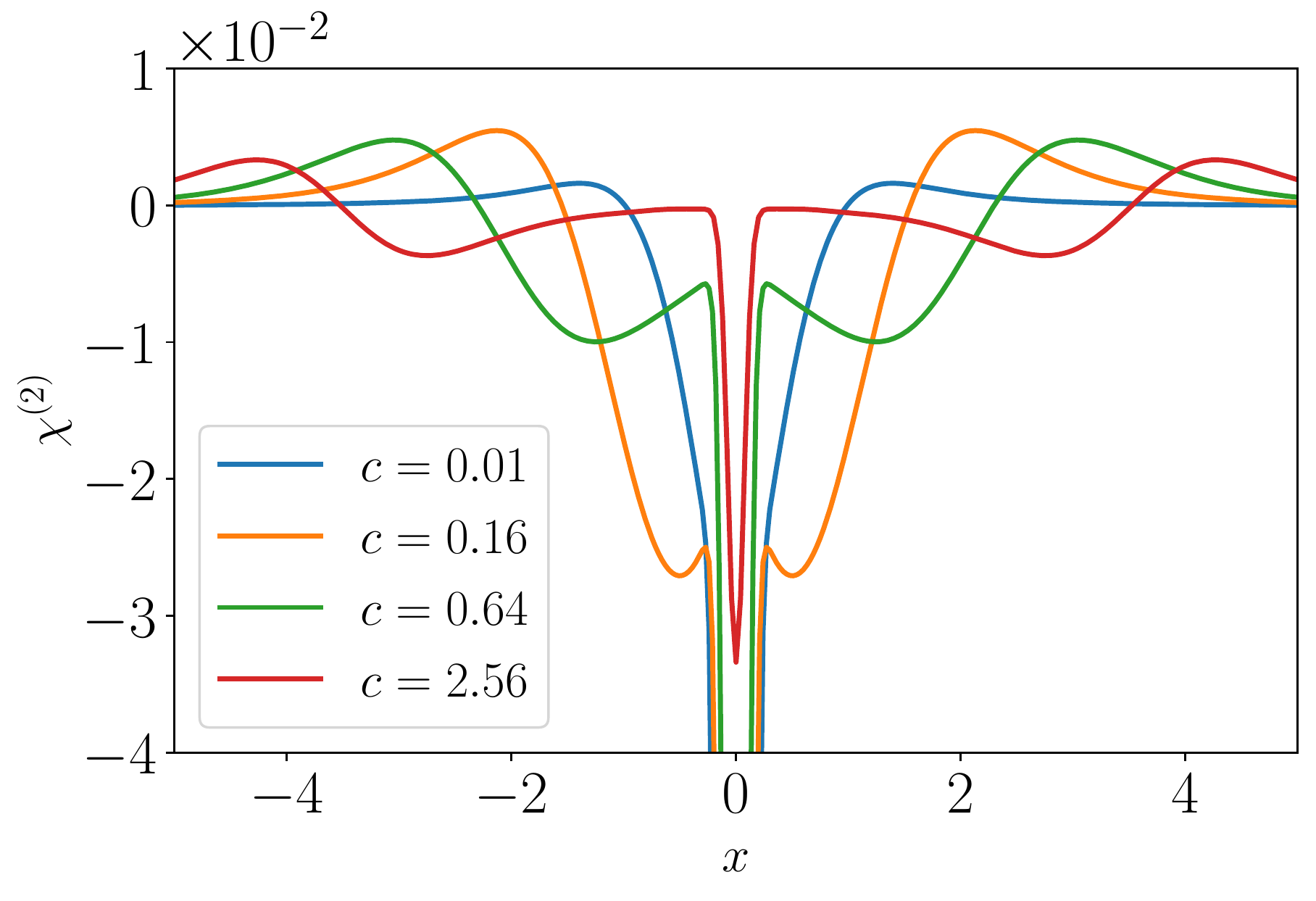}
			\caption{$\chi^{(2)}(0, 0; x, \tau; 0, \tau+t)$ in the Lieb-Liniger model for various interaction strengths $c$.
				We take $T = 2$, $\bar{n}=1$, $\tau=t=1$ and regularize the $\delta$-function GHD propagator  as a Gaussian of width $\eta= 0.1$. In a noninteracting system, the only response would come from the resolution-limited spike at $x_1 = 0$; everything else is a signature of interactions.}
			\label{fig:chi2-plot}
		\end{figure}

		Indeed, this response is readily computed using the above formalism (as detailed in the Methods section and~\cite{SM}); Fig.~\ref{fig:chi2-plot} shows the results for various interaction strengths, at fixed temperature $T$ and boson density $\bar{n}$. As $c$ decreases we see that the signal moves closer to $x=0$. This is because for $c\to0$ the system is proximate to  a Bose-Einstein condensate at $c=0$ and $T=0$ (see e.g. Refs.~\cite{takahashi_1999, JiangYu-Zhu:50311}), and hence  only slow, low momentum quasiparticle states are occupied. [See the Methods for another effect contributing to the signal moving near $x=0$.] Furthermore, note that the signal starts to decrease either for $c\lesssim 10^{-2}$ or $c\gtrsim 1$, as expected. [Recovering free boson response as $c\to 0$ requires studying very low $c$; as $c$ decreases, the density of states initially {\it increases} due to the incipient Bose condensation,  enhancing  interaction effects.] These  observations are not restricted to the protocol analyzed above: any  protocol which  separates the same-ray `free' contribution from the regular part of the response would yield similar results. Thus, nonlinear correlators provide a more direct window into the interacting Lieb-Liniger gas than linear response.
		
		In passing, note that spatially-resolved measurements of multi-point nonlinear response would also give a powerful diagnostic for ballistic transport, and hence integrability. As we remarked above, the existence of nontrivial Euler-scale response---absent strict Galilean invariance---is a hallmark of integrable dynamics, and suffices to diagnose integrability. Even in Galilean-invariant chaotic fluids with a few conserved currents, quasiparticles propagate sub-ballistically, so one expects an Euler-scale multi-point correlator like that shown in Fig.~\ref{fig:chi2-plot} to be strongly suppressed relative to the integrable case.
		
	\section{Higher-order Drude weights}
	
		So far, we have focused on  spatially-resolved response. 
		While this can be measured in cold-atom experiments, most solid-state spectroscopic techniques only access spatially integrated quantities. At the Euler scale, the most natural
		integrated quantity is the generation of a persistent current in response to a uniform electric field. This follows from the fact that the current operator in an integrable system generically has some part that is strictly conserved under time evolution, so the current generated in response to an electric field will not decay over time. For example, specializing to first-order response, $\int dx\, \chi^{(1)}_{\hat{j}_0}(0,0;x,t)$ will tend to a constant as $t\to\infty$; this limiting value is called the Drude weight~\cite{Bertini_transport_review,prosen_drude}.
		Alternatively, in frequency space, the conductivity goes as $\sigma(\omega) = \pi \mathcal{D} \delta(\omega) + \ldots$.
		Drude weights extend to nonlinear response: a field $E$ applied to the system for a finite time $\Delta t$ drives a persistent current $j_0(\varphi)$, where $\varphi \equiv E \Delta t$ is the vector potential variation due to the field.
		By expanding $j_0(\varphi)$ as a series in its argument and taking derivatives, we may define a sequence of nonlinear Drude weights~\cite{PhysRevB.102.165137,Watanabe2020} (which can be defined similarly for any other operator).
		
		Our diagrammatic approach can  straightforwardly be used to compute $N^{\text{th}}$ order Drude weights $\mathcal{D}^{(N)}$, by integrating the $N^{\text{th}}$ order response over the positions of the field insertions. As shown in the SM~\cite{SM}, this yields the recursive formula
		\begin{equation}\label{eq:DrudeRecursion}
		\mathcal{D}^{(N)}_{\hat{O}} = -\int d\theta_{N} a^\eff \partial_{\theta_N} n \frac{\delta}{\delta n_{\theta_{N}}} \mathcal{D}^{(N-1)}_{\hat{O}},
		\end{equation}
		with $\mathcal{D}^{(0)}= \langle \hat{O} \rangle$. This recursive formula allows to obtain a closed-form expression for non-linear Drude weight of arbitrary order only using TBA technology, with explicit expressions up to third order given in the SM~\cite{SM}.
		
		While \eqref{eq:DrudeRecursion} rapidly becomes  complex with increasing $N$, a simple limit emerges for the first term of a  high-temperature expansion: since each factor of $\partial_\theta n$ is proportional to $T^{-1}$, the leading contribution to $\mathcal{D}^{(N)}_{\hat{O}}$ is always obtained by acting with $\frac{\delta}{\delta n_{\theta_{N}}}$ on the factor $\partial_{\theta_{N-1}} n_{\theta_{N-1}}$  in $\mathcal{D}^{(N-1)}_{\hat{O}}$. Integrating by parts, we find that as $T
		\to\infty$,
		\begin{equation}
		\mathcal{D}^{(N)}_{\hat{O}} = -\int d\theta \partial_\theta n \left[ a^\eff \frac{\partial}{\partial\theta}\right]^{N-1}a^\eff \frac{\delta \langle \hat{O} \rangle}{\delta n(\theta)}+O(T^{-2}).
		\end{equation}
		We benchmark this GHD result against numerical simulations of a paradigmatic integrable model, the XXZ spin chain, and focus on spin current response. Since spatial inversion symmetry forces spatially-averaged current response functions to vanish for any even $N$,  we focus on $\mathcal{D}^{(3)}$. We work in the easy-axis limit, and exploit the generalized  Kohn formula~\cite{PhysRevB.102.165137,Watanabe2020} combined with exact diagonalization (ED) on small systems. (Unfortunately, state-of-the-art matrix product operator techniques for linear Drude weights~\cite{PhysRevLett.108.227206} do not give comparably good results for higher-order Drude weights~\cite{SM}.) Our results are presented in Fig.~\ref{fig:xxznumerics}; despite the difficulty of extrapolating reliably to the thermodynamic limit from the small system sizes accessible to ED, we see that the GHD results are within the range of our extrapolations at high temperature, and agree extremely well at lower temperatures.  We also see good agreement as we vary the easy-axis anisotropy at fixed temperature. 
		
	\section{Discussion}
	
		In this work we have presented a general framework for computing nonlinear response within GHD,  demonstrated that it is in excellent agreement with exact numerics, and  illustrated how it can directly distinguish  between free and interacting integrable systems. Our results suggest a natural experimental protocol for directly measuring quasiparticle interaction effects in the Lieb-Liniger model using ultracold atomic gases. (Importantly, this approach does not require single-site imaging resolution.) Since our proposal involves \emph{finite-time} behavior, it can be applied to realistic experimental settings where integrability is only approximate. We have focused on regimes where the nonlinear response is perturbative, and can be expanded in powers of the field strength. In such  regimes, our results for nonlinear response bear some resemblance to those 
		for full counting statistics~\cite{Doyon2019, 10.21468/SciPostPhys.8.1.007, 2020arXiv201206496P}. The multipoint correlators that appear in that theory (with all operators evaluated at the same point in space) are a special case of those computed here.
		
		We need not look far for integrable systems in which response is inherently \emph{nonperturbative}. The most transparent example is the isotropic Heisenberg model, at $h = 0$. In linear response, this model exhibits anomalous transport in the Kardar-Parisi-Zhang universality class~\cite{Ljubotina_nature, Ilievski18, GV19, NMKI19, Ljubotina19, Vir20,NGIV20}. We may approach this regime from  nonzero $\beta h$ by taking appropriate limits.
		Explicitly computing the spin current due to an impulse $\varphi = E \Delta t$, we find that
		\begin{equation}
		J(h, \varphi) \approx h \sum_{s = 1}^{1/h} s^{-4} f(h \varphi s^3),
		\end{equation}
		for some scaling function $f$ that is approximately sinusoidal in its argument~\cite{SM}. The sum is over quasiparticle ``strings'', which are bound states of $s$ elementary magnons. If we now take $\varphi \to 0$ at fixed $h \neq 0$, we obtain  a series in powers of $\varphi$, where the first term is the linear Drude weight ($\varphi \mathcal{D}^{(1)}\sim \varphi h^2 |\log h|$), the next nonvanishing term is $\mathcal\varphi^3 \mathcal{D}^{(3)} \sim \varphi^3 / h^2$, and higher-order terms are even more singular in the half-filling limit. The $\varphi \to 0$ and $h \to 0$ limits strikingly fail to commute: if we instead take $h \to 0$ at fixed $\varphi$, we find  that $J(h, \varphi) \sim h^2 \varphi |\log h \varphi|$. In effect, $\varphi$ can act as a \emph{cutoff} on response: for any fixed field, sufficiently large bound states respond nonperturbatively and undergo Bloch oscillations.  
		A proper description of such nonperturbative phenomena requires extending the present framework beyond Euler scale, e.g., by including diffusive corrections~\cite{DeNardis2018, Gopalakrishnan18, DeNardis_SciPost} and other sources of irreversibility~\cite{PhysRevLett.122.240606}. We leave this as an important direction for future work.
		
		\textit{Note added.---} As this paper was being completed we became aware of recent work~\cite{2021arXiv210305838T} that computes exact non-linear Drude weights for the XXZ chain. Ref.~\cite{2021arXiv210305838T}  considers only $T=0$ and $|\Delta|<1$, and hence has limited overlap with the results presented here. We have checked that our results for $T\to 0$ agree in the relevant regime of $\Delta$. Since the issue of irreversibility for finite-$T$ GHD calculations is particularly challenging to address in the easy-plane regime for reasons noted in Ref.~\cite{PhysRevLett.122.240606}, we defer detailed study of this regime to future work.
		
		\section*{Acknowledgements}
		
		We thank Bruno Bertini for insightful discussions. We acknowledge support from NSF Grant No. DMR-1653271 (S.G.),  the European Research Council under the European Union Horizon 2020 Research and Innovation Programme via Grant Agreement No. 804213-TMCS (S.A.P., S.B.),  the US Department of Energy, Office of Science, Basic Energy Sciences, under Early Career Award No. DE-SC0019168 (R.V.),  and the Alfred P. Sloan Foundation through a Sloan Research Fellowship (R.V.).
		
	\section{Materials and Methods}
	
		\subsection*{Computation of $\Gamma^{(2)}$}
		
		In this subsection, we describe how $\Gamma^{(2)}= \frac{\delta^2 n_\theta(z)}{\delta n_{\theta_1}(z_1) \delta n_{\theta_2}(z_2)}$ can be expressed in terms of linear propagators $D$ and a scattering vertex. For the most general case of $\Gamma^{(p)}$, we refer the reader to the SM~\cite{SM}.
		
		To compute $\Gamma^{(2)}$  we take the functional derivative of \eqref{eq:homogeneous-GHD} w.r.t. $n(x_0, t_0)$ and $n(x_1,t_1)$, and evaluate it on top of a homogeneous background, obtaining
		\begin{multline}
		\left(\partial_t + v^\eff_\theta \partial_x\right) \Gamma^{(2)} =\\
		=- \left( \int d\theta'\,\frac{\delta v^\eff_\theta}{\delta n_{\theta'}} D_{\theta',\theta_1}(z_1,z) \partial_x D_{\theta,\theta_0}(z_0,z) + (0 \leftrightarrow 1) \right).
		\end{multline}
		Note that, since we have now fixed $\vec{n}$ to be the uniforrm thermal  background, we have dropped terms proportional to $\partial_x n$.
		The LHS of this equation consists of  $\Gamma^{(2)}$ acted upon by a linear partial differential operator (PDO) [since $\vec{n}$ is now fixed to be the thermal background] whose Green's function is given by the propagator $D$. Inverting the PDO using its Green's function, we have
		\begin{multline}
		\Gamma^{(2)} = - \delta_{\theta,\theta_2}\int d^2 z_s\, D_{\theta}(z_s,z) \frac{\delta v^\eff_\theta}{\delta n_{\theta_1}} D_{\theta_1}(z_1,z) \partial_x D_{\theta}(z_0,z)+ \\+ (0 \leftrightarrow 1)
		\end{multline}
		where we introduced $D_\theta(z_0,z_1)= \delta(x_1-x_0-v^\eff_\theta (t_1-t_0))$, $z_s=(x_s,t_s)$ labelling the position of the scattering process, and $d^2z_s= dx_s\,dt_s$.
		
		In this expression we can recognize the structure of a process like that depicted in Fig.~\ref{fig:diagrams-chi2}d.
		Note that $\frac{\delta v^\eff_\theta}{\delta n_{\theta_1}}$ and hence $
		\Gamma^{(2)}$ will be non-zero only if the model is interacting; in a free theory $v^\eff$ reduces to the group velocity and will hence be  independent of $n_{\theta_1}$.
		
		\subsection*{$\chi^{(2)}$ in the Lieb-Liniger model}
		
		In this section we focus on the protocol described in the Section ``Measuring interactions in the Lieb-Liniger gas'' and the corresponding computation of $\chi^{(2)}_{\hat{q}_0}(0,0;x,\tau; 0,\tau+t)$. In particular, for $x\neq 0$, $\chi^{(2)}$ is given by the sum of two contributions, represented in Fig.~\ref{fig:diagrams-chi2}(c-d). In fact, (a) is zero whenever $x\neq0$, and (b) is zero in the Lieb-Liniger model since $a^\eff=1$ and does not carry any dependence on the state $n$.
		
		For continuity with the previous section, we focus on contribution (d), which is given by
		\begin{equation}
		\chi^{(2)}_d = \int d\theta\, d\theta_1\, d\theta_2\,  a^\eff_{\theta_1} \partial_\theta n_{\theta_1} a^\eff_{\theta_2} \partial_\theta n_{\theta_2} 
		\Gamma^{(2)} \frac{\delta \langle \hat{O} \rangle}{ \delta n(\theta)},
		\end{equation}
		where $\Gamma^{(2)}= \frac{\delta n_\theta (0,t+\tau)}{\delta n_{\theta_1}(0,0) \delta n_{\theta_2}(x,\tau)}$ is given in the previous section in terms of $\frac{\delta v^\eff_\theta}{\delta n_{\theta_1}}$. In the Lieb-Liniger model the momentum corresponds to the rapidity $k=\theta$ and the energy is given by $e = k^2/2$ [as customary, we are choosing units in which the mass of the particles is $1$]. The bare group velocity is then given by $v_\theta=k=\theta$. The effective group velocity, which is renormalized by the interactions is then given by the solution of the integral equation~\cite{Doyon_notes}
		\begin{equation}
		\label{eq:MM-veff}
		\rho^t_\theta v^\eff_\theta = \rho^t_\theta v_\theta + \int {d\theta'}\, K_{\theta-\theta'}  n_{\theta'} \rho^t_{\theta'} v^\eff_{\theta'}.
		\end{equation}
		$K_{\theta-\theta'}$ is the so-called scattering kernel, which encodes  the phase shifts (or equivalently time delays) of quasiparticles upon scattering. In the Lieb-Liniger model it takes the form
		\begin{equation}
		K_{\theta-\theta'} = \frac{1}{\pi}\frac{c}{(\theta-\theta')^2+c^2}.
		\end{equation}
		
		Before separately analysing the two limits $c\to0$ and $c\to\infty$, we report the free particle result, which holds both for free fermions or bosons, and is entirely due to diagram (a):
		\begin{equation}
		\chi^{(2)}_a = \int \frac{dp}{2\pi} \frac{\delta \langle \hat O \rangle}{\delta n_p} D_p(z_2,z_1) a_p \partial_p \left( D_p(z_1,z_0) a_p \partial_p n_p \right).
		\end{equation}
		As previously noted, the products $D_p(z_2,z_1) D_p(z_1,z_0)$ and $D_p(z_2,z_1) \partial_p D_p(z_1,z_0)$ vanishes whenever all the points $\{z_0,z_1,z_2\}$ do not lie on the same ray. Finally, we can see that the only difference between fermions and bosons is in the dependence of $n_p$, i.e.
		\begin{equation}
		n_p = \frac{1}{1\pm e^{\beta(e_p-\mu)}},
		\end{equation}
		in the two cases.
		
		For the Lieb-Liniger has, it is easiest to recover this form in the free-fermion limit $c\to\infty$, in which $K_{\theta-\theta'}\to0$. In this case, it is then clear that $v^\eff_p \to v_p = p$ independently of the state $n_\theta$. As a consequence $\frac{\delta v^\eff_\theta}{\delta n_{\theta_1}}\to0$ and $\Gamma^{(2)}$ will vanish.

		The free-boson limit $c\to0$ of the Lieb-Liniger gas is more subtle. The key observation is that the width of the function $K_{\theta-\theta'}$ is proportional to $c$. Combining this observation with Eq.~\eqref{eq:MM-veff} we expect that $\frac{\delta v^\eff_\theta}{\delta n_{\theta_1}}$ will be non-negligible only if $\theta-\theta_1\lesssim c$. Looking at Fig.~\ref{fig:diagrams-chi2}(d), note that the slope of the black trajectory is given by $v^\eff(\theta_1)$, while the slope of the blue one is $v^\eff(\theta)$. Thus, as $c\to0$, for an effective scattering process to take place $v^\eff(\theta)-v^\eff(\theta_1)=O(c)$, requiring that the three points lie approximately on the same ray, i.e. $x=O(c)$. We can then see that ultimately this contribution will be peaked in the same region where diagram (a) is non-zero and it will be impossible to separate them. Similar considerations would also hold for diagram (c).
		
		While the above discussion implies that $\tilde{\chi}^{(2)}(x\neq0,t,\tau)$ tends to zero in the $c\to0$ limit, as it should for a free-particle system, it is not immediately clear analytically that the signal at $x=0$ tends to its free boson value. This can, however, be verified numerically, by showing that the sum of diagrams (c), and (d) in Fig.~\ref{fig:diagrams-chi2} tends to zero as $c\to0$.

		\subsection*{Numerical computation of the non-linear Drude weights}
		
		In our numerical calculations we used the generalized Kohn formula~\cite{PhysRevB.102.165137,Watanabe2020} combined with exact diagonalization. The generalized Kohn formula relates the current Drude weights to the derivatives of the energy levels when a gauge flux $\varphi$ is threaded through a system with periodic boundary conditions. E.g. for $\mathcal{D}^{(3)}_{\hat{j}_0}$ it gives
		\begin{equation}
		\mathcal{D}^{(3)}_{\hat{j_0}} = \frac{1}{L} \sum_n p_n \frac{d^4\epsilon_n}{d\varphi^4} =  \frac{1}{L} \sum_n p_n \frac{d^3\langle \hat{J}_0 \rangle_n}{d\varphi^3},
		\end{equation}
		where $L$ denotes the length of the system, $n$ runs over the eigenstates of $\hat{H}_0$, each of whom has energy $\epsilon_n$ and is occupied with probability $p_n$. In the second part $\hat{J}_0$ is the total charge current $\sum_j \hat{j}_0(j)$ and $\langle \cdot \rangle_n$ denotes the average over the $n$-th eigenstate. The figures reported in the main text are obtained by summing over all symmetry sectors (momentum and magnetization).
		
		Note that a naive implementation of this formula based on finite differences would be problematic. For small enough $\varphi$ the numerical precision on the finite difference (which must the be divided by $\varphi^3$) would limit the accuracy of the results. On the other hand, at large enough $\varphi$, level crossings start to occur, thus compromising the results. Empirically, it seems that these two problems significantly compromise the results for all values of $\varphi$ starting at $L\gtrsim 15$.
		There are two possible solutions to this problem. One is to use perturbation theory to express $\frac{d^4\epsilon_n}{d\varphi^4}$ based on matrix elements of $\hat{H}_0$ and $\hat{J}_0$ (see Eq.~(31) of Ref.~\cite{Watanabe2020}). Another alternative exploits the integrability of the model in question. In fact, we could choose a large $\varphi\simeq 10^{-2}$, and track levels through the various crossings based on their fidelity $\braket{n(\varphi_0)|n(\varphi_1)}$. Both approaches give consistent results for the cases we considered.
		
		Finally, we point out that this approach is heavily limited by finite-size effects, specifically at small $|\Delta|-1$ or medium-high temperatures, where a reliable extrapolation to the thermodynamic limit is not possible~\cite{SM}.
	
%

	
\newpage


\section*{Supplementary material}

# SUPPLEMENTARY MATERIAL FOR "Hydrodynamic non-linear response of interacting integrable systems"

Michele Fava,[1] Sounak Biswas,[1] Sarang Gopalakrishnan,[2] Romain Vasseur,[3] and S. A. Parameswaran[1]

[1]*Rudolf Peierls Centre for Theoretical Physics, Clarendon Laboratory, Oxford OX1 3PU, UK*
[2]*Department of Physics, The Pennsylvania State University, University Park, PA 16802, USA*
[3]*Department of Physics, University of Massachusetts, Amherst, MA 01003, USA*
(Dated: October 3, 2021)

## CONTENTS

**Appendix A: Diagrammatic rules**

In this section we explain and derive the diagrammatic rules mentioned in the main text. We first explain the formalism, and then devote our attention to (i) deriving the two distinct classes of field vertices described in the main text, and (ii) proving that the scattering vertices $\Gamma^{(p)}$ can be constructed by considering connected diagrams formed by propagators and scattering vertices. Finally, we explain how to extend the formalism to compute the response to potentials coupling to higher charges $\hat{q}_j$.

As observed in the main text, the starting point is to consider the evolution of an initially-thermal state under the Hamiltonian

$$\hat{H}(t) = \hat{H}_0 + \sum_j \int dx\, V_j(x,t)\hat{q}_j(x). \tag{A1}$$

where we have generalized our discussion to encompass generic conserved charges of the integrable model (rather than restricting to a global $U(1)$ charge as in the main text). In this case, Eq. (2) of the main text generalizes to[1–7]

$$\partial_t n(\theta) + v^{\text{eff}}(\theta)\partial_x n(\theta) = -\sum_j \left[V_j v_j^{\text{eff}}(\theta)\partial_x n(\theta) + E_j a_j^{\text{eff}}(\theta)\partial_\theta n(\theta)\right], \tag{A2}$$

where $E_j = -\partial_x V_j$, $a_j^{\text{eff}} = q_j^{\text{dr}}/(k')^{\text{dr}}$ is the effective acceleration produced by $E_j$, and $v_j^{\text{eff}} = (q_j')^{\text{dr}}/(k')^{\text{dr}}$ is the change in the effective velocity produced by $V_j$. [Note that for clarity in this Supplement we have changed how we label rapidities, eschewing the $n_\theta$ notation of the main text in favour of the more explicit $n(\theta)$.]

As explained in the main text, the starting point of our calculation is the factorization of Eq. (3), reproduced here for clarity:

$$\chi_{\hat{O}}^{(N)}(\{x_n, t_n\}) = \prod_{n=0}^{N-1} \frac{\delta}{\delta E(x_n, t_n)} \langle \hat{O}(x_N, t_N) \rangle \bigg|_{E \to 0} \tag{A3}$$

using Eq. (4), again reproduced here:

$$\frac{\delta \langle O(z_1) \rangle}{\delta E(z_0)} = \int d\theta d\alpha \frac{\delta \langle O(z_1) \rangle}{\delta n_\alpha(z_1)} \frac{\delta n_\alpha(z_1)}{\delta n_\theta(z_0)} \frac{\delta n_\theta(z_0)}{\delta E(z_0)}. \tag{A4}$$

Combining these results allows us to identify the following elementary objects that appear in any response calculation:

- the **propagator** evaluated on the thermal background $D(\theta, \theta'; z, z')$, describing the linearized advection of quasiparticles

- the **non-linear generalization of the propagator** $\Gamma^{(p)}$ with $p > 1$, describing multi-particle processes which modify the linear advection given by $D(\theta, \theta'; z, z')$

- the **field vertices**, corresponding $\delta n(\theta, x, t+0^+)/\delta E_j(x, t)$ and its higher, functional derivatives w.r.t. $\boldsymbol{n}(\cdot, x, t)$. These reflect the perturbation in the normal modes produced by the field $E_j$. As noted in the main text, field vertices come in two varieties, depending on whether they excite a quasiparticle from rest, or accelerate a previously-excited quasiparticle.

- the **measurement vertices**, i.e. the expectation value $\langle \hat{O}(x,t) \rangle$ and the functional derivatives w.r.t. $\boldsymbol{n}(x,t)$.

- in general, we must also consider **potential vertices** for perturbations that couple to charges with $j > 0$, since in such cases we must also account for the variation $\delta n(\theta, x, t+0^+)/\delta V_j(x, t)$ due to the effect of $v_j^{\text{eff}}$ in the RHS of Eq. (A2). This contribution is absent for $j = 0$, since $U(1)$ gauge invariance forbids any direct dependence on the potential $V_j$ rather than the field $E_j$.

Crucially, each vertex only involves quantities within the same space-time cell, so that it can be computed within the TBA formalism. The only object that is non-trivial to compute is the non-linear generalization of the propagator $\Gamma^{(p)}$. As we will show in Sec. A 2, $\Gamma^{(p)}$ can be expressed in terms of the linear propagator $D(\theta, \theta'; z, z')$ and a new class of vertices, namely

- the **scattering vertices**, corresponding to functional derivatives of $v^{\text{eff}}[\boldsymbol{n}(x,t)](\theta)$ w.r.t. $\boldsymbol{n}(x,t)$.

With the above definitions, the calculation of $\chi^{(N)}$ can be represented diagrammatically as follows. Each vertex corresponds to a vertex in the diagram. The set of possible vertices is reported in table in Table SI, together with a cartoon of the corresponding process in the same schematic language used in Fig. 2 in the main text. Note that field and scattering vertices have a 'marked' incoming leg. In order to diagrammatically indicate the distinction between unmarked and marked legs, we place them respectively to the left and right of a 'notch' on these vertices; therefore, at most one line can enter the vertex to the right of the notch. For a field vertex, there can either be no incoming line to the right of the notch, corresponding to the case when a field accelerates a quasiparticle previously at rest, or a single incoming line, corresponding to the case where the field accelerates a quasiparticle that was already disturbed from the thermal distribution at a previous time. The remaining 'unmarked' incoming lines to the left of the notch describe any other excited quasiparticles in the same cell that modify the effective acceleration induced by the field. A scattering vertex always has an incoming line on the right of the notch, representing the quasiparticle whose effective velocity is modified by scattering at the spacetime cell due to its scattering against the other excited quasiparticles in the cell, which are represented by the lines entering the unmarked legs to the left of the notch. Diagrams are then formed by joining the vertices reported in the table with lines, corresponding to linearized propagators. Finally, $\chi^{(N)}$ is given by the sum of all connected diagrams that can be constructed with one measurement vertex and with $F$ field vertices and $P$ potential vertices, with the constraint $F + P = N$. To fully describe the diagrammatic calculation we must then specify the rules for how to associate a real number to each diagram.

Before reporting these rules, we stress that at any given $N$ only a finite number of diagrams can contribute. This follows from two facts: each field vertex can increase the number of propagators *at most* by 1, while all other vertices *decrease* the number of propagators. Consequently one can see that that any scattering or measurement vertex that

can enter a diagram contributing to $\chi^{(N)}$ can have at most $N$ incoming legs, while a field vertex contributing to such a diagram can have at most $N-1$ incoming legs.

The recursive procedure for computing the weight of a given diagram to do this may be summarized as follows:

1. Label the rapidity of each line. Assign different rapidity labels to different lines, unless the rapidity is constrained to be the same according to the graphical rules in Table SI. This happens when a line enters the rightmost/marked leg of a field, scattering, or potential node, in which case the outgoing line needs to have the same rapidity of the line entering the marked leg.

2. Recursively evaluate of the weight of each [sub-]diagram using the following procedure:

    (a) Identify the uppermost object in the diagram, i.e. the one corresponding to the latest time. In the first recursive step this will be the measurement vertex; in later steps it can be any other vertex or a propagator.

    (b) Remove this object. This will split the diagram into one ore more connected sub-diagrams.

    (c) Evaluate the weight of each connected sub-diagram generated by the removal in the previous step.

    (d) The weight of the sub-diagram under consideration is given by the product of two terms: (i) the expression $I_1$ of the removed object (as in Table SI) and (ii) a second term $I_2$, coming from the weight of the sub-diagrams, as detailed below

      - If the removed object is a propagator, $I_2$ is the weight of the remaining sub-diagram
      - If the removed object is either a measurement node or a field node without lines entering the marked leg (as on the second row of Table SI), $I_2$ is the product of the weights of the remaining sub-diagrams.
      - If the removed object is a field node with one line entering the marked leg (as the on the third row of Table SI), $I_2$ is the product of the weights of the sub-diagrams ending in the non-marked leg -i.e. the one denoted by $\theta_1,\ldots,\theta_k$ in the table- times the derivative w.r.t. $\alpha$ of the weight of the sub-diagram entering the marked leg.
      - If the removed object is a scattering vertex (fifth row) or a potential vertex (sixth row), $I_2$ is the product of the weights of the sub-diagrams ending in the unmarked legs —i.e., those denoted by $\theta_1,\ldots,\theta_k$ in the table — multiplied by the spatial derivative of the weight of the sub-diagram entering the marked leg w.r.t. the position of the removed vertex. Operationally, this simply replaces the propagator $\delta\left(x - x' - v^{\text{eff}}(\alpha)(t-t')\right)$ entering the marked leg, with its derivative $\delta'\left(x - x' - v^{\text{eff}}(\alpha)(t-t')\right)$.

3. Integrate over each rapidity label.

4. Integrate over the space and time coordinates of all scattering vertices.

As an example we report the diagrams contributing to $\chi^{(2)}$ in Fig. S2. For an example on how to evaluate these and how to compute the first few functional derivatives, see Sec. D. Evidently, the diagrammatic formalism is simply a schematic representation of the physical processes in Fig. 2 of the main text, but in a form that is more conveniently transcribed into a concrete calculation, and is hence more suited to systematic computation of higher-order responses. This example illustrates how the diagrammatic formalism can be systematically extended to higher $N$.

Before proceeding to the derivation of the field and scattering vertices, we comment on the case where $V_j \neq 0$ for $j > 0$. Setting for concreteness $N = 2$, the response to a certain perturbation will be given by

$$\delta\langle \hat{O} \rangle = \frac{\delta^2 \langle \hat{O} \rangle}{\delta E_j(z_0) \delta E_j(z_1)} E_j(z_0) E_j(z_1) + \frac{\delta^2 \langle \hat{O} \rangle}{\delta V_j(z_0) \delta E_j(z_1)} V_j(z_0) E_j(z_1) + $$
$$+ \frac{\delta^2 \langle \hat{O} \rangle}{\delta E_j(z_0) \delta V_j(z_1)} E_j(z_0) V_j(z_1) + \frac{\delta^2 \langle \hat{O} \rangle}{\delta V_j(z_0) \delta V_j(z_1)} V_j(z_0) V_j(z_1). \quad (A5)$$

In general, $\delta\langle \hat{O} \rangle$ will be given by the sum of $2^N$ distinct expressions -some of which are zero. Each of these corresponds to a different class of diagram, depending on the $N$ choices — one at each perturbed point $z_0, \ldots, z_{N-1}$ — where we may introduce a field vertex or a potential vertex. In other words, the choices of potential or field vertex determines which functional derivative is being computed, i.e. every field vertex introduces a $\frac{\delta}{\delta E_j}$ and every potential vertex a $\frac{\delta}{\delta V_j}$.

In the next two subsections, we first derive the form of the field vertices reported in the main text and in Table SI of the field vertex, and demonstrate that $\Gamma^{(p)}$ is given by the sum over all possible connected diagrams composed of propagators and scattering vertices, and with $n$ incoming lines and an outgoing one.

TABLE SI. List of all vertices composing a non-linear response diagram. The "Cartoon" depicts the process in the same style used in the main text, while the "Diagram" column reports a more schematic representation, which is the one used throughout in the Supplementary material and is used in actual diagrammatic calculations— see e.g. Fig S1 and S2. In all vertices the space-time position of the $\delta n$ perturbations coincide with the position of the vertex. The two version of the field vertex reflect the possibility of evaluating $\partial_\alpha n$ of the thermal background, i.e. accelerating a previously thermal quasiparticle, or take the functional derivative of $\partial_\alpha n$ with one incoming line, i.e. accelerating a quasiparticle already acted upon by the field (see also Fig. 2). The scattering vertex has a similar bipartite structure, where $k > 1$ incoming legs act with a functional derivative on $v^{\text{eff}}$, while one acts on $\partial_{x'} n(\alpha, x', t')$, where $z' = (x', t')$ is the space-time position of the vertex and must be integrated upon.

| **Process** | **Cartoon** | **Diagram** | **Expression** |
|---|---|---|---|
| Linearized propagation, $D(\theta; z, z')$ | | | $\theta_H(t-t')\delta\left(x - x' - v^{\text{eff}}(\theta)(t-t')\right)$ |
| Field accelerates thermal quasiparticle | | | $-\dfrac{\delta^k a^{\text{eff}}(\alpha)}{\delta n(\theta_1)\cdots\delta n(\theta_k)}\partial_\alpha n(\alpha),\ k \geq 0$ |
| Field accelerates excited quasiparticle | | | $-\dfrac{\delta^k a^{\text{eff}}(\alpha)}{\delta n(\theta_1)\cdots\delta n(\theta_k)},\ k \geq 0$ |
| Measurement | | | $\dfrac{\delta^k \langle \hat{O}\rangle}{\delta n(\theta_1)\cdots\delta n(\theta_k)},\ k \geq 1$ |
| Scattering | | | $-\dfrac{\delta^k v^{\text{eff}}(\alpha)}{\delta n(\theta_1)\cdots\delta n(\theta_k)},\ k \geq 1$ |
| Potential modifies velocity of quasiparticle | | | $-\dfrac{\delta^k v_j^{\text{eff}}(\alpha)}{\delta n(\theta_1)\cdots\delta n(\theta_k)},\ k \geq 0$ |

### 1. The field vertex

In this subsection we derive the explicit expression for the field vertex. Specifically, we willll show that the perturbation $\delta n$ in the occupation factors produced by the application of the field $V_j$ is given by

$$\delta n(\theta_0, x_0, t_0 + 0^+) = -a_j^{\text{eff}}[\boldsymbol{n}_0(x_0, t_0)](\theta)\partial_\theta n_0(\theta_0, x_0, t_0)\delta E_j(x_0, t_0) - v_j^{\text{eff}}[\boldsymbol{n}_0(x_0, t_0)](\theta)\partial_x n_0(\theta_0, x_0, t)\delta V_j(x_0, t_0), \quad \text{(A6)}$$

at linear order in $\delta E_j$ and $\delta V_j$, but for a generic initial state (i.e. background occupation) $n_0(\theta; x, t)$. This will then allow us to compute non-linear responses as detailed in the main text, by taking $n_0$ to be a non-stationary state produced by the field perturbations acting at earlier times.

The starting point is the GHD equation (A2). Since, without loss of generality, we are restricting to the computation of the linear term in the field, we can simply evaluate the right hand side of (A2) directly on $n_0$. We are thus left with an equation of the form

$$\partial_t n(\theta; x, t) + v^{\text{eff}}(\theta)[n(x,t)]\partial_x n(\theta; x, t) = J(\theta; x, t)[\boldsymbol{n}_0(x,t)], \quad \text{(A7)}$$

where the source term $J$ is determined by $n_0$

$$J(\theta; x, t)[\boldsymbol{n}_0(x,t)] = -\sum_j \left[ V_j v_j^{\text{eff}}(\theta) \partial_x n_0(\theta) + E_j a_j^{\text{eff}}(\theta) \partial_\theta n_0(\theta) \right]. \tag{A8}$$

We thus define a retarded Green's function

$$G(\theta, x, t; \theta_0, x_0, t_0)[\boldsymbol{n}_0] = \frac{\delta n(\theta, x, t)}{\delta J(\theta_0, x_0, t_0)}, \tag{A9}$$

with the boundary condition $n(x, t < t_0) = n_0(x, t)$. As usual the Green's function determines $\delta n$ by

$$\delta n(\theta, x, t_0 + 0^+) = \int d\theta_0 \int dx_0 \, G(\theta, x, t_0 + 0^+; \theta_0, x_0, t_0)[\boldsymbol{n}_0] \delta J(\theta_0, x_0, t_0)[\boldsymbol{n}_0]. \tag{A10}$$

The rest of this subsection is devoted to showing that under the assumption (whose validity we discuss following the derivation) that $n_0(\theta, x, t)$ is smooth w.r.t. $x$, the Green's function is given by

$$G(\theta, x, t_0 + 0^+; \theta_0, x_0, t_0) = \delta(\theta - \theta_0)\delta(x - x_0), \tag{A11}$$

whence Eq. (A6) follows immediately.

To prove Eq. (A11), we recognize that $G$ satisfies the equation

$$\underbrace{\partial_t G(\theta; x, t) + v^{\text{eff}}[n_0(x,t)](\theta) \partial_x G(\theta; x, t)}_{=A} + \underbrace{\int d\alpha \frac{\delta v^{\text{eff}}(\theta)}{\delta n(\alpha)}[n_0(x,t)] G(\alpha; x, t) \partial_x n_0(\theta; x, t)}_{=B} = \delta(x - x_0)\delta(t - t_0)\delta(\theta - \theta_0). \tag{A12}$$

While solving for the full space-time dependence of $G$ is generally hard (although, see Ref. 8 for a possible approach), to determine the field vertex we only need $G(x, t_0 + 0^+)$, i.e. in the instant immediately after the perturbation has acted. For this purpose, we integrate both sides of the equation inside a window in rapidity and space centred at $(\theta_0, x_0)$ and with space and rapidity width $2\Delta\theta$ and $2\Delta x$ respectively. In doing so, we keep in mind that the ultimate purpose is to take the limit of $(\Delta\theta, \Delta x) \to \boldsymbol{0}$. The integral of the RHS of Eq. (A12) is immediate and gives $\delta(t - t_0)$. We now separately analyze the two terms on the LHS of Eq. (A12) labeled $A$, and $B$.

First, we can recognize that $A$ can be rewritten as

$$A = \partial_t G(\theta; x, t) + v^{\text{eff}}[n_0(x,t)](\theta) \partial_x G(\theta; x, t) = \frac{d}{dt} G(\theta; x, t) \Big|_{(\tilde{x}(\theta,t), t)}. \tag{A13}$$

The last term represents the total derivative of $G$ evaluated along a characteristic trajectory -the space-time trajectory along which $n_0(\theta)$ is constant- with $\frac{d}{dt}\tilde{x}(\theta, t) = v^{\text{eff}}(\theta)$. This means that, if $B$ can be neglected, as we will show later, we must have

$$\int_{\theta_0 - \Delta\theta}^{\theta_0 + \Delta\theta} d\theta \int_{x_0 - \Delta x}^{x_0 + \Delta x} dx \frac{d}{dt} G(\theta; x, t) \Big|_{(\tilde{x}(\theta,t), t)} + O(\Delta\theta) = \delta(t - t_0). \tag{A14}$$

Taking the limit $(\Delta\theta, \Delta x) \to \boldsymbol{0}$, this readily implies Eq. (A11).

Finally, we show that the solution Eq. (A11) is consistent with the assumption that the integral of $B$ is of order $\Delta\theta \Delta x$. Assuming that $n_0$ is smooth and that the dependence of $v^{\text{eff}}$ on $n$ is also smooth, we have

$$\int_{\theta_0 - \Delta\theta}^{\theta_0 + \Delta\theta} d\theta \int_{x_0 - \Delta x}^{x_0 + \Delta x} dx \, B = 2\Delta\theta \int d\alpha \frac{\delta v^{\text{eff}}(\theta_0)}{\delta n(\alpha)}[n_0(x_0, t_0)] \partial_x n_0(\theta_0; x_0, t_0) \int_{x_0 - \Delta x}^{x_0 + \Delta x} dx \, G(\alpha, x, t) = O(\Delta\theta), \tag{A15}$$

thus completing the proof of Eq. (A11).

We close this subsection by commenting on the smoothness assumption of $n_0$ w.r.t. $x$ that we have used to show that the integral of $B$ is $O(\Delta\theta)$ and could thus be neglected. If $n_0(x)$ is not differentiable, then Eq. (A11) is not the solution of Eq. (A12); extra terms would appear in Eq. (A11), depending on the specific nature of the discontinuity/singularity in $n_0$. However, if we start from a homogeneous state, such singularities can only appear if (i) the time-evolution is evaluated *strictly* at the Euler scale and (ii) there is a time when the perturbing potential is not smooth. However, (i) and (ii) are ultimately inconsistent with each other. In fact if (ii) applies, the gradient expansion that underlies a hydrodynamic treatment is ultimately unjustified. [This is for example clear from the derivation of Eq. (A2) in Ref. 1,

which requires higher spatial derivatives of $V_j(x)$ to be suppressed. While this is ensured by the Euler scaling limit if $V_j(x)$ is smooth, this assumption breaks down if $V_j(x)$ has discontinuities.] On the other hand, we are ultimately interested in determining the response to localized perturbations in order to define $\chi^{(N)}(\{x_j, t_y\})$. We can resolve this tension by *defining* $\chi^{(N)}(\{x_j, t_y\})$ as the response to smooth fields whose amplitude is localized around position $x_j$. This can be achieved e.g. by considering Gaussian perturbations of spatial width $\eta$ (see e.g. Eq. (D10)). In this way, the Euler-scale treatment is justified, $n(x)$ will be a smooth function of spacetime, and Eq. (A11) will hold. Finally, we can take the limit $\eta \to 0$ at the end of the calculation. This will ultimately amount to regulating the calculation by introducing a Gaussian broadening of the propagator $D_\theta = \delta_\eta(x - v_\theta^{\text{eff}} t)$ with $\delta_\eta$ denoting a Gaussian of with $\eta$, and then recovering Euler-scale results by taking the regulator to zero.

### 2. The scattering vertex

In this section we will show show that adding the scattering vertex to the diagrammatic formulation properly accounts for the functional derivatives of the propagator, i.e. quantities of the form

$$\Gamma^{(p)} = \frac{\delta^p n(\alpha; x_p, t_p)}{\delta n(\theta_{p-1}; x_{p-1}, t_{p-1}) \cdots \delta n(\theta_0; x_0, t_0)} = \frac{\delta^p n(\alpha; x_p, t_p)}{\delta n_{p-1} \cdots \delta n_0}, \tag{A16}$$

where we have introduced a shorthand notation $n_i = n(\theta_i; x_i, t_i)$. In particular, we claim that $\Gamma^{(p)}$ can be obtained by summing all diagrams containing propagator lines and a single scattering vertex only, under the condition that the number of incoming lines is $p$ and there is only one outgoing line (or equivalently that the diagram is connected).

The general argument is a slight generalization of the case $p = 2$ reported in the main text and Methods section, which we recapitulate here. To compute $\Gamma^{(2)}$ we take the functional derivative of Eq. (A2) w.r.t. $n(x_0, t_0)$ and $n(x_1, t_1)$, with $V_j \equiv 0$ and evaluate it on top of a homogeneous background, obtaining

$$(\partial_t + v^{\text{eff}}(\alpha)\partial_x) \frac{\delta n(\alpha; y, \tau)}{\delta n_1 \delta n_0} = -\left( \int d\theta' \frac{\delta v^{\text{eff}}(\alpha)}{\delta n(\theta')} \frac{\delta n(\theta'; y, \tau)}{\delta n_1} \partial_y \frac{\delta n(\alpha; y, \tau)}{\delta n_0} + (0 \leftrightarrow 1) \right). \tag{A17}$$

On the RHS, we recognize the structure of a scattering vertex with two incoming legs, and the two $D$ propagators connected to it. The LHS of this equation consists of the required quantity $\Gamma^{(2)}$ acted upon by a linear partial differential operator [since $n$ is now fixed to be the thermal background] whose Green function is given by $D(\theta, \alpha, z_p, z_s)$ with $z_s = (y, \tau)$, corresponding to a $\delta n$-line exiting the scattering vertex. Inverting the PDO using its Green's function, we see that $\Gamma^{(2)}$ is indeed given by the sum of the two possible diagrams of the required form.

We will now proceed to prove the general statement by induction on $p$, with the (trivial) base case $\Gamma^{(1)} = D$. To prove the inductive step, we proceed in the same way as our computation of $\Gamma^{(2)}$ above. We begin with the GHD equation (A2) with $V_j \equiv 0$ and take its the functional derivatives w.r.t. the set $\{n_0, \ldots, n_{p-1}\}$

$$\partial_{t_p} \Gamma^{(n)}(\alpha; x_p, t_p) + \frac{\delta^n \left[ v^{\text{eff}}(\alpha) \partial_x n(\alpha; x_p, t_p) \right]}{\delta n_{p-1} \cdots \delta p_0} = 0, \tag{A18}$$

where we have left implicit the arguments $\{\theta_j, x_j, t_j\}$ of $\Gamma^{(p)}$. Upon evaluating the above equation on a static background $n$, this can be immediately recast as an equation for $\Gamma^{(p)}$

$$(\partial_t + v^{\text{eff}}(\alpha)\partial_x) \Gamma^{(p)}(\alpha; x_p, t_p) =$$
$$= -\sum_{I \in P} \frac{\delta^{(k)} v^{\text{eff}}(\alpha)}{\delta n(\bar{\theta}_1) \cdots \delta n(\bar{\theta}_k)} \frac{\delta^{\#I_1} n(\bar{\theta}_1; x_p, t_p)}{\prod_{j \in I_1} \delta n_j} \cdots \frac{\delta^{\#I_k} n(\bar{\theta}_k; x_p, t_p)}{\prod_{j \in I_k} \delta n_j} \partial_x \frac{\delta^{\#R_I} n(\alpha; x_p, t_p)}{\prod_{j \in R_I} \delta n_j}. \tag{A19}$$

In the above expression $P$ is a collection of sets $I = \{I_1, \cdots, I_k\}$ with the following properties: (i) each $I_j$ is a non-empty subset of $\{0, \cdots, p-1\}$, (ii) the $I_j$ are pairwise disjoint, (iii) $R_I = \{0, \cdots, p-1\} \setminus \bigcup_j I_j$ is a non-empty set. This reflects the fact that we are computing $\Gamma^{(p)}$ on top of a homogeneous state, i.e. $\partial_x n = 0$.

The solution of Eq. (A19) can be found as for the $p = 2$ case above. The LHS is a partial differential operator with a known Green's function acting on the quantity of interest. Therefore, denoting the 'source' term on the RHS by $J(\alpha; y, \tau) = \sum_{I \in P} J_I(\alpha; y, \tau)$, we find that

$$\Gamma^p(\alpha; x_p, t_p) = \int d\theta \int dy \int d\tau \, D(\alpha, \theta, z_p, (y, \tau)) J(\alpha; y, \tau) =$$
$$= \int d\theta \int dy \int d\tau \, D(\alpha, \theta, z_p, (y, \tau)) \sum_I J_I(\alpha; x', t') = \sum_I \Gamma_I^p(\alpha; x_p, t_p). \tag{A20}$$

We recognize in Eq. (A20) the structure of the scattering vertex reported in Table SI, situated in position $(y, \tau)$, and with $k$ legs entering the vertex from the left and one — corresponding to $R$ — from the right. Each of the incoming legs $\frac{\delta^{\#I_m} n(\bar{\theta}_m; y, \tau)}{\prod_{j \in I_m} \delta n_j} = \Gamma^{(\#I_m)}$. We then note that $\#I_m < n \; \forall m$ and $\#R_I < n$, so that the inductive hypothesis guarantees that $\forall m$ there is a set of diagrams $D_{I_k}$ whose sum gives $\Gamma^{(\#I_k)}$. Then, summing over the set of diagrams $D_{I_1} \times \cdots \times D_{I_k} \times D_{R_I}$ combined with the scattering vertex in $(y, \tau)$, we obtain $\Gamma_I^{(p)}$. Summing over $I \in P$, we then have that $\Gamma^{(n)}$ is given by a sum of connected diagrams composed of propagators and scattering vertices.

So far we have shown that $\Gamma^{(p)}$ is a sum of connected diagrams of propagators and scattering vertices. To complete the proof of our statement we need to show that each diagram $D$ of this type contributes to $\Gamma^{(p)}$ with a prefactor 1. To do this, we start by identifying in $D$ the last scattering vertex time-wise (which coincides with the vertex from which the last line emerges). Defining $k+1$ to be the number of incoming legs, the diagram in question can only contribute to $\Gamma_I^{(p)}$ for $\#I = k$. To determine the specific partition $I$, we remove the last scattering vertex. Due to the tree-like structure of the diagram, this operation splits the diagram into $k+1$ disconnected pieces $D_j$ ($0 \le j \le k-1$) - connected to the left side of the scattering vertex - and $D_R$ - connected to the right side. This automatically induces a partition $\tilde{I} = I \cup \{R\}$ on the legs associated to $\delta n_i$ ($0 < j < p-1$), depending on the $D_j$ to which they belong. We claim that $D$ is a diagram contributing to $\Gamma_I^{(n)}$. In fact, by the inductive hypothesis, each $D_j$ (and $D_R$) will be a diagram contributing to $\frac{\delta^{\#I_j} n(\bar{\theta}_j; x', t')}{\prod_{i \in I_j} \delta n_i}$. Therefore, putting back the last scattering vertex, we recognize that $D$ is a term contributing to $\Gamma_I^{(p)}$. Finally, note that the choices of $I$ and $R$ were forced by the structure of the diagram, so that the same diagram could not contribute to a $\Gamma_{I'}$ with $I' \neq I$. Thus the prefactor of $D$ is just given by the products of the prefactors of $D_j$ and $D_R$, which are 1 by the inductive hypothesis, thus completing the proof of our statement.

### Appendix B: Diagrammatic formalism for $n$-point charge correlators

The formalism presented in the main text and in the preceding section can also be easily adapted to the computation of $n$-point connected correlation functions of charges,

$$C(\{i_j, x_j, t_j\}_{1 \le j \le n}) = \langle \hat{q}_{i_1}(x_1, t_1) \cdots \hat{q}_{i_n}(x_n, t_n) \rangle_c, \tag{B1}$$

with $t_1 < t_2 < \cdots < t_n$. [Note that, as explained in Ref. 1 and 8 the operators can be always rearranged in this way since the commutators are sub-leading at the Euler scale.]

The starting point of the computation is the observation (reported in Ref. 8) that the correlator can be computed recursively through

$$C(\{i_j, x_j, t_j\}_{1 \le j \le n}) = \frac{\delta}{\delta \beta_{i_1}(x_1, t_1)} C(\{i_j, x_j, t_j\}_{1 \le j \le n}). \tag{B2}$$

The above equation must be interpreted as follows. The state of the system at a given time can be described by $n(\theta, x)$ as we have done so far, or equivalently through the spatially-dependent Lagrange multipliers $\beta_i(x)$ characterizing the local GGE that approximates the state on mesoscopic scales, viz. $\hat{\rho}(x) \sim \exp(-\sum_i \beta_i(x) \hat{q}_i(x))$.

The $\delta \beta_{i_1}(x_1, t_1)$ perturbation can be shown to be associated to a $\delta n$ perturbation through[8–10]

$$\frac{\delta n(\theta)}{\delta \beta_i} = -n(\theta) \left[1 - n(\theta)\right] q_i^{\rm dr}(\theta). \tag{B3}$$

From this point the calculation is then formally identical to the one described in the main text, with the only difference that $\frac{\delta n}{\delta E}$ is replaced by $\frac{\delta n}{\delta \beta_i}$. This observation allows us to use the same diagrammatic formalism of the main text also to compute charge correlators by replacing the $k$-legs field vertex with a $k$-leg "$i$-th charge" vertex, corresponding to

$$-\frac{\delta^k \left[ \int d\alpha\, n(\alpha)\left((1 - n(\alpha)) q_i^{\rm dr}(\alpha)\right]\right.}{\delta n(\theta_1) \cdots \delta n(\theta_k)} \tag{B4}$$

Finally, we note that the computation of many-point time-ordered correlators in GHD[8,11–13] is also sometimes referred to as determining 'non-linear response'. However, such time-ordered correlators are physically different from the causal non-linear response to sequential perturbations computed in this manuscript. The latter involves differences of correlators ordered on the Keldysh contour (see e.g. (F2)). While for linear response these can always be reduced to a single time-order correlator using the Kubo-Martin-Schwinger relations, this no longer holds for non-linear

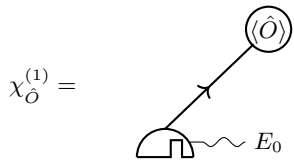

$$\chi_{\hat{O}}^{(1)} =$$

FIG. S1. Diagramatic expression for $\chi_{\hat{O}}^{(1)}$.

response[14–16]. Computing non-linear responses as defined in this work directly in terms of differences of correlators within the GHD formalism would be challlenging since at the Euler scale the $n$-point functions are independent of the ordering as noted in[8]. Therefore, computing the expectation value of the commutators by adding correlators would require to expand these beyond the leading Euler-scale results which are usually computed[8,11–13]. On the other hand, our approach focuses directly on perturbations of the Hamiltonian through inhomogeneous fields, thus overcoming the problem.

### Appendix C: Recovering linear response

In this section we show that the formalism described above correctly captures standard linear response functions at the Euler scale[9,10].

Linear response functions in GHD are normally computed by combining the fluctuation-dissipation theorem (see e.g. Refs. 17 and 18) with the formula for the connected two-point functions[9,10]. Focusing on the current response to an electric field — the discussion of the other cases is very similar — these considerations give the expression

$$\chi_{\hat{j}_i}^{(1)}(x,t) = \beta \int \frac{d\theta}{2\pi} (\partial_\theta k)^{\text{dr}}(\theta) n(\theta) \left[1 - n(\theta)\right] \left[v^{\text{eff}}(\theta)\right]^2 q_0^{\text{dr}}(\theta) q_i^{\text{dr}}(\theta) \delta\left(x - v^{\text{eff}}(\theta)t\right). \quad \text{(C1)}$$

Within our formalism linear response functions are immediately obtained from Eq. (4) in the main text, which we report here for the reader's convenience

$$\chi_{\hat{j}_i}^{(1)}(x,t) = \frac{\delta \langle \hat{j}_i(x,t) \rangle}{\delta E(0,0)} = \int d\theta d\alpha \frac{\delta \langle \hat{j}_i(x,t) \rangle}{\delta n(\alpha,x,t)} \frac{\delta n(\alpha,x,t)}{\delta n(\theta,0,0)} \frac{\delta n(\theta,0,0)}{\delta E(0,0)}, \quad \text{(C2)}$$

which would correspond to the diagram in Fig. S1 We now recall (i) the expression for the propagator $D(\alpha, \theta, (x,t), (0,0)) = \frac{\delta n(\alpha,x,t)}{\delta n(\theta,0,0)}$ (see e.g. Table SI), (ii) the expression for the field vertex $\frac{\delta n(\theta,0,0)}{\delta E(0,0)} = -a^{\text{eff}}(\theta) \partial_\theta n(\theta)$, where in a unperturbed thermal state

$$\partial_\theta n(\theta) = -\beta n(\theta) \left[1 - n(\theta)\right] (\partial_\theta e)^{\text{dr}}(\theta), \quad \text{(C3)}$$

and (iii) that[9,10]

$$\frac{\delta \langle \hat{j}_i \rangle}{\delta n(\alpha)} = \frac{1}{2\pi} (k')^{\text{dr}}(\alpha) v^{\text{eff}}(\alpha) q_i^{\text{dr}}(\alpha). \quad \text{(C4)}$$

Plugging these into Eq. (3), integrating over $\alpha$ using $\delta(\theta - \alpha)$, and rewriting $a^{\text{eff}}(\partial_\theta e)^{\text{dr}}$ as $v^{\text{eff}} q_0^{\text{dr}}$, we recover the result in Eq. (C1).

### Appendix D: Explicit expressions for the diagrams contributing to $\chi^{(2)}$

The four diagrams reported in Fig. S2, respectively give the following explicit expressions contributing to $\chi_{\hat{O}}^{(2)}(x_0, 0; x_1, \tau_0; x_2, \tau_1 + \tau_0)$. Before reporting their full expression we exemplify how we can use the evaluation scheme described in Sec. A to write the expression corresponding to a given diagram.

$\chi_1^{(2)}$: We start from the measurement node, which gives us a factor $\frac{\delta \langle \hat{O} \rangle}{\delta n(\theta)}$. The uppermost element in the remaining diagram is a propagator $D_\theta(x_2 - x_1, \tau_1)$. Removing the propagator, we find a field node without any leg entering from the left, which yields $a^{\text{eff}}(\theta)$. Next, according to the rules previously outlined, we need to multiply the result

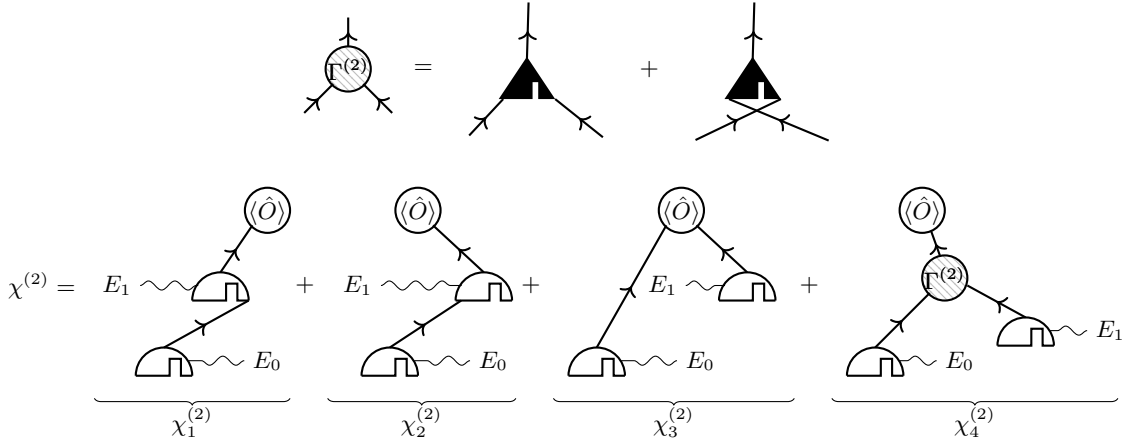

FIG. S2. Diagrams contributing to $\Gamma^{(2)}$ and $\chi^{(2)}$.

thus far with the derivative w.r.t. $\theta$ of the subdiagram that enters the marked leg of the field node. Proceeding similarly to before, we thus find $\partial_\theta \left( D_\theta(x_1 - x_0, \tau_0) a^{\text{eff}}(\theta) \partial_\theta n \right)$. Putting everything together we obtain

$$\chi_1^{(2)} = \int d\theta \left[ \frac{\delta \langle \hat{O} \rangle}{\delta n(\theta)} \right] \delta \left( x_2 - x_1 - v^{\text{eff}}(\theta) \tau_1 \right) a^{\text{eff}}(\theta) \partial_\theta \left\{ \delta \left( x_1 - x_0 - v^{\text{eff}}(\theta) \tau_0 \right) a^{\text{eff}}(\theta) \partial_\theta n, \right\} \tag{D1}$$

$\chi_2^{(2)}$: The evaluation of $\chi_2^{(2)}$ proceeds similarly to that of $\chi_1^{(2)}$, up to the uppermost field node (excluded). This first part yields $\frac{\delta \langle \hat{O} \rangle}{\delta n(\theta_2)} D_{\theta_2}(x_2 - x_1, \tau_1)$. The field node now has one leg connected to the left side, corresponding to a factor $\frac{\delta a^{\text{eff}}(\theta_2)}{\delta n(\theta_1)} \partial_{\theta_2} n(\theta_2)$ — note that the rapidity of legs entering a node from the left and the outgoing rapidity from that node need to be labelled by distinct variables. This has to be multiplied by the subdiagram connected to the incoming leg, which, as before, gives us $D_{\theta_1}(x_1 - x_0, \tau_0) a^{\text{eff}}(\theta_1) \partial_{\theta_1} n(\theta_1)$. Multiplying everything gives

$$\chi_2^{(2)} = \int d\theta_1 \left[ a^{\text{eff}}(\theta_1) \partial_\theta n(\theta_1) \right] \delta \left( x_1 - x_0 - v^{\text{eff}}(\theta_1) \tau_0 \right) \times$$
$$\times \int d\theta_2 \left[ \frac{\delta a^{\text{eff}}(\theta_2)}{\delta n(\theta_1)} \partial_\theta n(\theta_2) \right] \left[ \frac{\delta \langle \hat{O} \rangle}{\delta n(\theta_2)} \right] \delta \left( x_2 - x_1 - v^{\text{eff}}(\theta_2) \tau_1 \right), \tag{D2}$$

$\chi_3^{(2)}$: The measurement vertex now has two incoming legs, whose rapidity we can label by $\theta_1$ and $\theta_2$. This vertex then gives $\frac{\delta^2 \langle \hat{O} \rangle}{\delta n(\theta_1) \delta n(\theta_2)}$. This has to be multiplied by the two disconnected diagrams obtained by removing the measurement vertex. These are $D_{\theta_2}(x_2 - x_1, \tau_1) a^{\text{eff}}(\theta_2) \partial_{\theta_2} n(\theta_2)$ and $D_{\theta_1}(x_2 - x_0, \tau_1 + \tau_0) a^{\text{eff}}(\theta_1) \partial_{\theta_1} n(\theta_1)$. Explicitly, this yields

$$\chi_3^{(2)} = \int d\theta_1 \int d\theta_2 \left[ a^{\text{eff}}(\theta_1) \partial_\theta n(\theta_1) \right] \left[ a^{\text{eff}}(\theta_2) \partial_\theta n(\theta_2) \right] \left[ \frac{\delta^2 \langle \hat{O} \rangle}{\delta n(\theta_1) \delta n(\theta_2)} \right] \times$$
$$\times \delta \left( x_2 - x_1 - v^{\text{eff}}(\theta_2) \tau_1 \right) \delta \left[ x_2 - x_0 - v^{\text{eff}}(\theta_1)(\tau_1 + \tau_0) \right], \tag{D3}$$

$\chi_4^{(2)}$: There are two diagrams corresponding to $\chi_4^{(2)}$, which can be obtained by replacing the graphical representation of $\chi_4^{(2)}$ in Fig. S2 with the two terms corresponding with $\Gamma^{(2)}$ in Fig. S2. For concreteness we focus on the first one here and report the complete expression for the sum later. We call the space-time location of the scattering vertex $(y, s + \tau_0)$. The measurement vertex gives as usual $\frac{\delta \langle \hat{O} \rangle}{\delta n(\theta_2)}$. The remaining diagram terminates with a propagator - associated to $D_{\theta_2}(x_2 - y, \tau_1 - s)$. The topmost object is then the scattering vertex. This corresponds to $\frac{\delta v^{\text{eff}}(\theta_2)}{\delta n(\theta_1)}$. Removing the scattering vertex splits the diagram in two connected sub-diagrams. For the specific diagram we are considering, the left-most one gives $D_{\theta_1}(y - x_0, s + \tau_0) a^{\text{eff}}(\theta_1) \partial_{\theta_1} n(\theta_1)$. The

rightmost part can be evaluated similarly, however we must remember to take its derivative w.r.t. $y$, since it enters the marked/rightmost leg of the scattering vertex: $\partial_y D_{\theta_2}(y - x_1, s) a^{\text{eff}}(\theta_2) \partial_{\theta_2} n(\theta_1)$. Summing with the other diagram in $\chi_4^{(2)}$ then gives

$$\chi_4^{(2)} = -\int d\theta_1 \int d\theta_2 \int dy \int_0^{\tau_1} ds \, [a^{\text{eff}}(\theta_1)\partial_\theta n(\theta_1)] \, [a^{\text{eff}}(\theta_2)\partial_\theta n(\theta_2)] \left[\frac{\delta v^{\text{eff}}(\theta_2)}{\delta n(\theta_1)}\right] \left[\frac{\delta \langle \hat{O} \rangle}{\delta n(\theta_2)}\right] \times$$
$$\times \delta\left(y - x_0 - v^{\text{eff}}(\theta_1)(s + \tau_0)\right) \delta'\left(y - x_1 - v^{\text{eff}}(\theta_2)s\right) \delta\left(x_2 - y - v^{\text{eff}}(\theta_2)(\tau_1 - s)\right) +$$
$$- \int d\theta_1 \int d\theta_2 \int dy \int_0^{\tau_1} ds \, [a^{\text{eff}}(\theta_1)\partial_\theta n(\theta_1)] \, [a^{\text{eff}}(\theta_2)\partial_\theta n(\theta_2)] \left[\frac{\delta v^{\text{eff}}(\theta_1)}{\delta n(\theta_2)}\right] \left[\frac{\delta \langle \hat{O} \rangle}{\delta n(\theta_1)}\right] \times$$
$$\times \delta\left(y - x_1 - v^{\text{eff}}(\theta_2)s\right) \delta'\left(y - x_0 - v^{\text{eff}}(\theta_1)(s + \tau_0)\right) \delta\left(x_2 - y - v^{\text{eff}}(\theta_1)(\tau_1 - s)\right). \quad \text{(D4)}$$

The functional derivatives reported above can be readily computed using the identity

$$\frac{\delta f^{\text{dr}}(\theta)}{\delta n(\theta_1)} = K^{\text{dr}}(\theta, \theta_1) f^{\text{dr}}(\theta_1), \quad \text{(D5)}$$

where $f$ is an arbitrary function of $\theta$. The field and scattering vertices can be computed b combining this identity with the definitions

$$a^{\text{eff}} = \frac{q_0^{\text{dr}}}{(\partial_\theta k)^{\text{dr}}} \quad \text{(D6)}$$

$$v^{\text{eff}} = \frac{(\partial_\theta e)^{\text{dr}}}{(\partial_\theta k)^{\text{dr}}}. \quad \text{(D7)}$$

Finally, the expression for the measurement vertex can be found in the case where $\hat{O}$ is a charge or current density, in which case the vertex with one incoming line is given by[9,10]

$$\frac{\delta \langle \hat{q}_i \rangle}{\delta n(\theta_1)} = \frac{1}{2\pi} (k')^{\text{dr}}(\theta_1) q_i^{\text{dr}}(\theta_1), \quad \text{(D8)}$$

$$\frac{\delta \langle \hat{j}_i \rangle}{\delta n(\theta_1)} = \frac{1}{2\pi} (e')^{\text{dr}}(\theta_1) q_i^{\text{dr}}(\theta_1). \quad \text{(D9)}$$

From these, measurement vertices with more incoming lines can be derived using Eq. (D5).

Fig. 3 in the main text has been obtained by numerically evaluating the expressions above and regularizing the Dirac-$\delta$ function as

$$\delta(x) \mapsto \delta_\eta(x) = \frac{1}{\sqrt{2\pi}\eta} \exp\left(-\frac{x^2}{2\eta^2}\right). \quad \text{(D10)}$$

For $\chi_4^{(2)}$ the integrals over $s$ and $y$ were first analytically evaluated through Mathematica and the simplified analytical expression was used to speed up the numerical computation.

### 1. Consistency Checks on $\chi^{(2)}$

In this section, we perform several consistency checks on the above expression for $\chi^{(2)}$. In particular we will check $\int dx_2 \chi_{\hat{O}}^{(2)}$ for $\hat{O} = \hat{q}$ or $\hat{O} = \hat{e}$, as these are cases where we can independently derive the response-function from general principles. As a first step, we report the integrals over $x_2$ of the three terms above, as computed after regularizing the Dirac-$\delta$ functions. In the case of the third term, the integration over $y$ and $s$ has also been carried out.

$$\int dx_2 \, \chi_1^{(2)} = -\int d\theta \, \delta_\eta\left(x_1 - x_0 - v^{\text{eff}}(\theta)\tau_0\right) a^{\text{eff}} \partial_\theta n \partial_\theta \left\{\left[\frac{\delta \langle \hat{O} \rangle}{\delta n(\theta)}\right] a^{\text{eff}}(\theta)\right\}, \quad \text{(D11)}$$

$$\int dx_2 \, \chi_2^{(2)} = \int d\theta_1 \, [a^{\text{eff}}(\theta_1)\partial_\theta n(\theta_1)] \, \delta_\eta\left(x_1 - x_0 - v^{\text{eff}}(\theta_1)\tau_0\right) \int d\theta_2 \left[\frac{\delta a^{\text{eff}}(\theta_2)}{\delta n(\theta_1)} \partial_\theta n(\theta_2)\right] \left[\frac{\delta \langle \hat{O} \rangle}{\delta n(\theta_2)}\right], \quad \text{(D12)}$$

$$\int dx_2\, \chi_3^{(2)} = \int d\theta_1 \int d\theta_2 \, [a^{\text{eff}}(\theta_1)\partial_\theta n(\theta_1)]\, [a^{\text{eff}}(\theta_2)\partial_\theta n(\theta_2)] \left[\frac{\delta^2\langle \hat{O}\rangle}{\delta n(\theta_1)\delta n(\theta_2)}\right] \times$$
$$\times \delta_{\sqrt{2}\eta}\left[x_1 - x_0 - v^{\text{eff}}(\theta_1)(\tau_1 + \tau_0) + v^{\text{eff}}(\theta_2)\tau_1\right], \quad \text{(D13)}$$

$$\int dx_2\, \chi_4^{(2)} = -\int d\theta_1 \int d\theta_2 \frac{[a^{\text{eff}}(\theta_1)\partial_\theta n(\theta_1)]\,[a^{\text{eff}}(\theta_2)\partial_\theta n(\theta_2)]}{v^{\text{eff}}(\theta_1) - v^{\text{eff}}(\theta_2)} \left[\frac{\delta v^{\text{eff}}(\theta_2)}{\delta n(\theta_1)} \frac{\delta\langle\hat{O}\rangle}{\delta n(\theta_2)} - \frac{\delta v^{\text{eff}}(\theta_1)}{\delta n(\theta_2)} \frac{\delta\langle\hat{O}\rangle}{\delta n(\theta_1)}\right] \times$$
$$\times \left[\delta_{\sqrt{2}\eta}\left(x_1 - x_0 - \tau_0 v^{\text{eff}}(\theta_1) - \tau_1(v^{\text{eff}}(\theta_1) - v^{\text{eff}}(\theta_2))\right) - \delta_{\sqrt{2}\eta}\left(x_1 - x_0 - \tau_0 v^{\text{eff}}(\theta_1)\right)\right]. \quad \text{(D14)}$$

Specializing to the case of $\hat{O} = \hat{q}_i$, the sum of the three terms simplifies to

$$\int dx_2 \chi_{\hat{q}_i}^{(2)}(x_0, 0; x_1, \tau_1; x_2, \tau_2) = \frac{1}{2\pi} \int d\theta_1 \left[a^{\text{eff}}(\theta_1)\partial_\theta n(\theta_1)\right] \delta(x_1 - x_0 - v^{\text{eff}}(\theta_1)\tau_0) \times$$
$$\times \left\{-\partial_\theta \left[q_0^{\text{dr}}(\theta_1)q_i^{\text{dr}}(\theta_1)\right] + \int d\theta_2 \partial_\theta n(\theta_2)\left[q^{\text{dr}}(\theta_2)K^{\text{dr}}(\theta_2, \theta_1)q_i^{\text{dr}}(\theta_1) + q_i^{\text{dr}}(\theta_2)K^{\text{dr}}(\theta_2, \theta_1)q_0^{\text{dr}}(\theta_1)\right]\right\}. \quad \text{(D15)}$$

Note that $\int dx_2 \chi^{(2)}$ does not depend on $\tau_1$, as it should be since $\int dx_2\, \hat{q}_i(x_2)$ commutes with the Hamiltonian $\hat{H}_0$. This can be further simplified by applying the identity

$$\partial_\theta h^{\text{dr}} = (\partial_\theta h)^{\text{dr}} + K^{\text{dr}}(\partial_\theta n)\, h^{\text{dr}}, \quad \text{(D16)}$$

on the term $\partial_\theta [q_0^{\text{dr}}(\theta_1) q_i^{\text{dr}}(\theta_1)]$. Thus we obtain

$$\int dx_2 \chi_i^{(2)}(x_0, 0; x_1, \tau_0; x_2, \tau_1 + \tau_0) = \frac{1}{2\pi}\int d\theta_1\, \left[a^{\text{eff}}(\theta_1)\partial_\theta n(\theta_1)\right] \delta(x_1 - x_0 - v^{\text{eff}}(\theta_1)\tau_0) \times$$
$$\times \left[(\partial_\theta q(\theta_1))^{\text{dr}}\, q_i^{\text{dr}}(\theta_1) + (\partial_\theta q_i(\theta_1))^{\text{dr}}\, q_0^{\text{dr}}(\theta_1)\right]. \quad \text{(D17)}$$

Finally, note that $\hat{q}$ is a charge associated to a global symmetry, and therefore $q(\theta)$ is constant, leaving us with the final expression

$$\int dx_2 \chi_i^{(2)}(x_0, 0; x_1, \tau_0; x_2, \tau_1 + \tau_0) = \frac{1}{2\pi}\int d\theta_1\, [a^{\text{eff}}(\theta_1)\partial_\theta n(\theta_1)]\, \delta(x_1 - x_0 - v^{\text{eff}}(\theta_1)\tau_0)\, (\partial_\theta q_i(\theta_1))^{\text{dr}}\, q_0^{\text{dr}}(\theta_1). \quad \text{(D18)}$$

Note that this result takes non-trivial contributions from all the diagrams discussed.

### a. Spatial integral of the charge response

The first check we perform is based on the following observation: if the Hamiltonian $\hat{H}_0$ in Eq. (A1) has a global symmetry, corresponding to a conserved charge $\hat{Q}_0 = \int dx\, \hat{q}_0(x)$, then the same charge $\hat{Q}$ is also conserved under the time evolution generated by (A1). It immediately follows that

$$0 = \frac{\delta^N \langle \hat{Q}_0 \rangle}{\delta E_0 \cdots \delta E_{N-1}} = \int dx_N\, \chi_{\hat{q}_0}^N. \quad \text{(D19)}$$

Specializing to the case $N = 2$, we see that the expression (D18) respects this condition, as can be verified by putting $q_i = q$ and using again that $q(\theta)$ is constant.

### b. Spatial integral of the energy response

Another case where we can independently compute the integral over $x_2$ of the response function is when $\hat{O}$ is the energy density $\hat{e}$. In this case, in fact, we are measuring the total energy increase produced by the application of the electric fields, i.e.

$$\chi_{\hat{H}_0}^{(2)}(x_0, 0; x_1, t_1) = \int dx_2\, \chi_{\hat{e}}^{(2)}(x_0, 0; x_1, t_1; x_2, t_2) = -\langle [\hat{E}(x_0, 0), [\hat{E}(x_1, t_1), \hat{H}_0]]\rangle, \quad \text{(D20)}$$

where $\hat{E}$ is an operator implementing the action of the electric field (see e.g. Eq. (F1)).

To compute this quantity we take the following route. First of all, we define the charge-charge response as

$$\chi^{(1)}_{\hat{q}_0,\hat{q}_0}(x_0,0;x_1,t_1) = +i\langle[\hat{q}_0(x_0,0),\hat{q}_0(x_1,t_1)]\rangle, \tag{D21}$$

where, as in the rest of this subsection, we include in the subscript not only the observable whose response is measured, but also the operators that act as perturbation. Next, we relate the $t_1$-derivative of the above expression to the second-order energy response

$$\partial_{t_1}\chi^{(1)}_{\hat{q}_0,\hat{q}_0}(x_0,0;x_1,t_1) = +i\langle\left[\hat{q}_0(x_0,0),i\left[\hat{H}_0,\hat{q}_0(x_1,t_1)\right]\right]\rangle = -\chi^{(2)}_{\hat{q}_0,\hat{q}_0,\hat{H}_0}(x_0,0;x_1,t_1). \tag{D22}$$

We now want to relate the left-hand side to the electric field-current response function $\chi_{\hat{E},\hat{j}}$, and the right-hand side to the energy response to electric fields $\chi_{\hat{E},\hat{E},\hat{H}_0}$. We start by discussing how to transform the response to a charge into the response to a field. Denoting with $\varphi(x)$ the electric potential associated with an electric field configuration $E = -\partial_x\varphi$, we have that the same response can be written either as

$$\int dx_j \chi_{\cdots,\hat{q}_0,\cdots,\hat{O}}(\cdots;x_j,t_j;\cdots)\varphi(x_j), \tag{D23}$$

or

$$\int dx_j \chi_{\cdots,\hat{E},\cdots,\hat{O}}(\cdots;x_j,t_j;\cdots)E(x_j). \tag{D24}$$

Integrating the first expression by parts and equating the integrands we find

$$\partial_{x_j}\chi_{\cdots,\hat{E},\cdots,\hat{O}}(\cdots;x_j,t_j;\cdots) = \chi_{\cdots,\hat{q}_0,\cdots,\hat{O}}(\cdots;x_j,t_j;\cdots). \tag{D25}$$

Applying this last formula on Eq. (D22), we thus obtain

$$\partial_{x_0}\partial_{t_1}\chi^{(1)}_{\hat{E},\hat{q}_0}(x_0,0;x_1,t_1) = -\partial_{x_0}\partial_{x_1}\chi^{(2)}_{\hat{E},\hat{E},\hat{H}_0}(x_0,0;x_1,t_1). \tag{D26}$$

Finally, we use the continuity equation for the charge on the left-hand side to obtain

$$\partial_{x_0}\partial_{x_1}\chi^{(1)}_{\hat{E},\hat{j}_0}(x_0,0;x_1,t_1) = \partial_{x_0}\partial_{x_1}\chi^{(2)}_{\hat{E},\hat{E},\hat{H}_0}(x_0,0;x_1,t_1). \tag{D27}$$

Noting that both response functions can only depend on $x_1 - x_0$ and, by causality, they tend to zero at $|x_1 - x_0| \to \infty$, we can directly equate the response functions. So, in the language of the main text, we have

$$\chi^{(1)}_{\hat{j}_0}(x_0,0;x_1,t_1) = \int dx_2\,\chi^{(2)}_{\hat{e}}(x_0,0;x_1,t_1;x_2,t_2). \tag{D28}$$

It is now easy to see that this identity is exactly satisfied by Eq. (D18) if we substitute $q_i(\theta)$ with $e(\theta)$.

#### c. Recovering free-fermion response

We now consider a free fermion system perturbed by the application of an electric field. We first directly compute the response function using fermionic anti-commutation relations e and then argue that the GHD approach recovers this result.

To directly compute the free-fermion response to an electric field, we consider first the density-density-density response. Defining the number operator as

$$\hat{n}(x,t) = \frac{1}{L}\sum_{k,k'}e^{i(k-k')x}e^{-i(\epsilon_k-\epsilon_{k'})t}\hat{c}^\dagger_{k'}\hat{c}_k, \tag{D29}$$

we compute

$$\chi^{(2)}_{\hat{n},\hat{n},\hat{n}}(x_0,0;x_1,\tau_0;0,\tau_1+\tau_0) = -\langle[\hat{n}(x_0,0),[\hat{n}(x_1,\tau_0),\hat{n}(0,\tau_1+\tau_0)]]\rangle. \tag{D30}$$

Using the fermionic anti-commutation relations we obtain

$$\chi^{(2)}_{\hat{n},\hat{n},\hat{n}}(x_0,0;x_1,\tau_0;0,\tau_1+\tau_0) = \frac{1}{L^3}\sum_{k,k',k''}[n(k')-n(k)]\,e^{i(k''-k')x_1}e^{i(k'-k)x_0}e^{-i(\omega_k-\omega_{k''})(\tau_1+\tau_0)}e^{-i(\omega_{k''}-\omega_{k'})\tau_0}+$$
$$+\frac{1}{L^3}\sum_{k,k',k''}[n(k')-n(k'')]\,e^{i(k'-k)x_1}e^{i(k''-k')x_0}e^{-i(\omega_k-\omega_{k''})(\tau_1+\tau_0)}e^{-i(\omega_{k'}-\omega_k)\tau_0}. \quad \text{(D31)}$$

Taking the Fourier transform, we arrive at

$$\chi^{(2)}_{\hat{n},\hat{n},\hat{n}}(k_0,k_1;\omega_0,\omega_1) = \frac{1}{L}\sum_k [n(k+k_0)-n(k)]\frac{i}{\omega_1-(\epsilon_k-\epsilon_{k+k_1+k_0})+i0^+}\frac{i}{\omega_0-(\epsilon_k-\epsilon_{k+k_0})+i0^+}+$$
$$-\frac{1}{L}\sum_k [n(k+k_0+k_1)-n(k+k_1)]\frac{i}{\omega_1-(\epsilon_k-\epsilon_{k+k_1+k_0})+i0^+}\frac{i}{\omega_0-(\epsilon_{k+k_1}-\epsilon_{k+k_0+k_1})+i0^+}. \quad \text{(D32)}$$

Euler-scale results are then retrieved by first sending $L\to\infty$, and then $k_0$, $k_1$, $\omega_1$, and $\omega_2$ to zero, while keeping their ratios fixed. We thus arrive at

$$\chi^{(2)}_{\hat{n},\hat{n},\hat{n}}(k_0,k_1;\omega_0,\omega_1) = k_0 k_1 \int \frac{dk}{2\pi}\frac{1}{\omega_1+(k_0+k_1)v(k)+i0^+}\partial_k\left[\frac{\partial_k n(k)}{\omega_0+k_0 v(k)+i0^+}\right]. \quad \text{(D33)}$$

Having the density-density-density response in momentum space, we can derive the usual field response through the identity

$$(ik_0)(ik_1)\chi_{\hat{n}}(k_0,k_1;\omega_0,\omega_1) = \chi^{(2)}_{\hat{n},\hat{n},\hat{n}}(k_0,k_1;\omega_0,\omega_1). \quad \text{(D34)}$$

Within GHD, we can instead compute the free-fermion response by making the dressing and effectivization transformation trivial. In this case the only non-vanishing contribution is given by $\chi^{(2)}_1$ in Eq. (D1). We can recognize its Fourier transform to be equal to the response function derived in this subsection using elementary methods.

### Appendix E: Non-linear Drude weights

The higher-order Drude weights can be computed as the long-time limit of the optical response

$$\mathcal{D}^{(N)}_{\hat{O}} = \lim_{\{\tau_j\to\infty\}}\int dx_0\cdots\int dx_{N-1}\chi^{(N)}(x_0,0;x_1,\tau_1;x_2,\tau_1+\tau_2;\cdots). \quad \text{(E1)}$$

When performing the integral over all spatial coordinates, we see that the diagrams that contain scattering vertices vanish, because they contain a spatial derivative of a propagator. Meanwhile, in all other diagrams the propagators integrate to 1. This can be easily seen as follows. Imagine integrating over $x_j$ in order of increasing $j$. In this way, when we need to integrate over $x_j$, there is a single propagator depending on $x_j$: the propagator of the $\delta n$ perturbation produced in $x_j$, while all other propagators entering $x_j$ have already been integrated. Then, if the propagator has not been differentiated, the integral givess 1; instead, if the propagator has been differentiated (i.e. it terminates in the 'marked leg' of a scattering vertex), then the integral is zero.

This observation enormously simplifies the recursive computation of the Drude weights. Since the Drude weights are just the product of nodes in the diagrams, we can recognize that the whole diagrammatic construction is just a book-keeping method to compute the recursive relation

$$\mathcal{D}^{(N)}_{\hat{O}} = -\int d\theta_N a^{\text{eff}}(\theta_N)\partial_{\theta_N}n(\theta_N)\frac{\delta}{\delta n(\theta_N)}\mathcal{D}^{(N-1)}_{\hat{O}}, \quad \text{(E2)}$$

with $\mathcal{D}^{(0)} = \langle\hat{O}\rangle$. This is Eq. (6) of the main text.

Using this formula, we can find closed-form expressions for the Drude weights recursively. For example, the second-order Drude weight associated to the (charge-)current $\hat{j}_0$ is given by

$$\mathcal{D}^{(2)}_{\hat{j}_0} = \int d\theta_1\, a^{\text{eff}}(\theta_1)\tilde{J}(\theta_1)\partial_{\theta_1}\left[a^{\text{eff}}(\theta_1)\partial_{\theta_1}n(\theta_1)\right] + \int d\theta_2\, a^{\text{eff}}(\theta_2)[\partial_{\theta_2}n(\theta_2)]\int d\theta_1\,\frac{\delta\left[a^{\text{eff}}(\theta_1)\tilde{J}(\theta_1)\right]}{\delta n(\theta_2)}\partial_{\theta_1}n(\theta_1), \quad \text{(E3)}$$

where

$$\tilde{J}(\theta) = \frac{\delta \langle \hat{j}_0 \rangle}{\delta n(\theta)} = \frac{1}{2\pi} \left( \partial_\theta e(\theta) \right)^{\mathrm{dr}} q^{\mathrm{dr}}(\theta). \tag{E4}$$

A much more compact expression can instead be found for the second-order Drude weight associated to a charge, just by integrating Eq. (D18) over $x_1$

$$\mathcal{D}^{(2)}_{\hat{q}_i} = -\frac{1}{2\pi} \int d\theta a^{\mathrm{eff}}(\theta) \partial_\theta n(\theta) q_0^{\mathrm{dr}}(\theta) \left[ \partial_\theta q_i(\theta) \right]^{\mathrm{dr}}. \tag{E5}$$

Finally, we report here the expression for the third-order Drude weight associated to a current

$$\begin{aligned}
\mathcal{D}^{(3)} = &- \int d\theta \, a^{\mathrm{eff}}(\theta) \tilde{J}(\theta) \partial_\theta \left[ a^{\mathrm{eff}}(\theta) \partial_\theta \left[ a^{\mathrm{eff}}(\theta) \partial_\theta n(\theta) \right] \right] + \\
&- 2 \int d\theta_2 \, a^{\mathrm{eff}}(\theta_2) \partial_{\theta_2} n(\theta_2) \int d\theta_1 \frac{\delta \left[ a^{\mathrm{eff}}(\theta_1) \tilde{J}(\theta_1) \right]}{\delta n(\theta_2)} \partial_{\theta_1} \left[ a^{\mathrm{eff}}(\theta_1) \partial_{\theta_1} n(\theta_1) \right] + \\
&- \int d\theta_2 \, a^{\mathrm{eff}}(\theta_2) \partial_{\theta_2} n(\theta_2) \int d\theta_1 \, a^{\mathrm{eff}}(\theta_1) \tilde{J}(\theta_1) \partial_{\theta_1} \left[ \frac{\delta a^{\mathrm{eff}}(\theta_1)}{\delta n(\theta_2)} \partial_{\theta_1} n(\theta_1) \right] + \\
&- \int d\theta_2 \, a^{\mathrm{eff}}(\theta_2) \partial_{\theta_2} \left[ a^{\mathrm{eff}}(\theta_2) \partial_{\theta_2} n(\theta_2) \right] \int d\theta_1 \frac{\delta \left[ a^{\mathrm{eff}}(\theta_1) \tilde{J}(\theta_1) \right]}{\delta n(\theta_2)} \partial_{\theta_1} n(\theta_1) + \\
&- \int d\theta_3 \, a^{\mathrm{eff}}(\theta_3) \partial_{\theta_3} n(\theta_3) \int d\theta_2 \, \partial_{\theta_2} n(\theta_2) \int d\theta_1 \, \partial_{\theta_1} n(\theta_1) \frac{\delta}{\delta n(\theta_3)} \left[ a^{\mathrm{eff}}(\theta_2) \frac{\delta \left[ a^{\mathrm{eff}}(\theta_1) \tilde{J}(\theta_1) \right]}{\delta n(\theta_2)} \right].
\end{aligned} \tag{E6}$$

While the formulae become increasingly complex with $N$, a simple limit is always obtained for the first term of a $\beta \to 0$ expansion. In fact, every factor $\partial_\theta n$ carries a factor of $\beta$, as discussed in the main text. Thus the leading-order term is always obtained by acting in Eq. (E2) with $\frac{\delta}{\delta n(\theta_N)}$ on the factor $\partial_{\theta_{N-1}} n(\theta_{N-1})$ appearing in $\mathcal{D}^{(N-1)}_{\hat{O}}$. Integrating by parts, we then obtain

$$\mathcal{D}^{(N)}_{\hat{O}}(\beta \to 0) = -\int d\theta a^{\mathrm{eff}} \partial_\theta n \partial_\theta \left[ a^{\mathrm{eff}} \partial \left[ \cdots \partial \left[ a^{\mathrm{eff}} \frac{\delta \langle \hat{O} \rangle}{\delta n(\theta)} \right] \right] \right] + O(\beta^2), \tag{E7}$$

with $\partial_\theta$ acting $N-1$ times, which is Eq. (7) of the main text.

### Appendix F: Numerical computation of the non-linear Drude weights

It is appealing to consider computing non-linear susceptibilities and non-linear Drude weights numerically, e.g. through standard MPO algorithms. For example in the XXZ chain, one could define the electric field operator as

$$\hat{E}(j_0) = -\sum_j \mathrm{sign}(j - j_0 + 1/2) \hat{S}^z_j. \tag{F1}$$

From this one could, e.g. compute the third order susceptibilities as

$$\chi^{(3)}_{\hat{O}}(x_0, t_0; x_1, t_1; x_2, t_2, x_3, t_3) = -i \mathrm{Tr} \left( \hat{\rho} \left[ \hat{E}(x_0, t_0), \left[ \hat{E}(x_1, t_1), \left[ \hat{E}(x_2, t_2), \hat{O}(x_3, t_3) \right] \right] \right] \right), \tag{F2}$$

where $\hat{O}(x_3)$ could be any local operator centered in position $x_3$ — for instance, the spin-current density — $\hat{\rho}$ is an equilibrium density matrix, and the time dependence is given by the Heisenberg representation w.r.t. $\hat{H}_0$. The non-linear Drude weight $\mathcal{D}^{(3)}_{\hat{O}}$ can then be obtained by integrating over three of the four spatial coordinates, and then taking the limit of $(t_1 - t_0, t_2 - t_1, t_3 - t_2) \to +\infty$. Note that, while $\hat{E}(j_0)$ is an extensive operator, its time-evolution will be non-trivial only in a light-cone region centered in $j_0$, thus making the above definition amenable to computations not affected by finite-size effects. For each value of the arguments $(t_1 - t_0, t_2 - t_1, t_3 - t_2)$, the integration over three spatial coordinates necessitates $L^3$ different contractions, each involving five MPOs (three operators from field

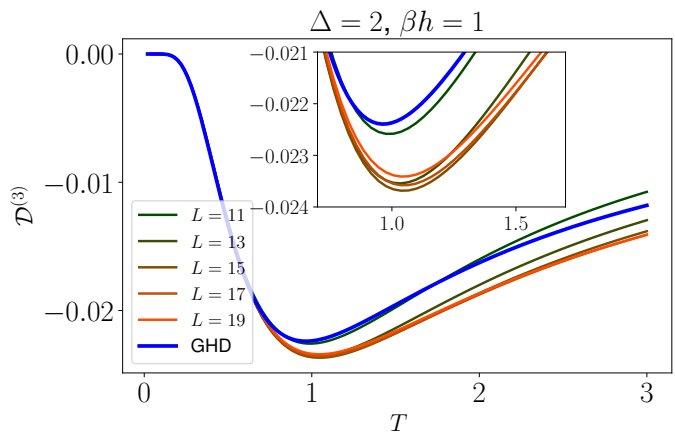

FIG. S3. Example of finite size effects at $\Delta = 2$ and at medium-high temperatures. The inset highlights the non-monotonic behaviour of the finite-size corrections.

insertions, one from the measurement, and the equilibrium density matrix). Each such contraction has a complexity of $O(Ld^3\chi^4\chi_\rho\chi_O)$ (where $\chi$ is the bond dimension of the time-evolved MPOs, $\chi_\rho$ is the bond dimension of the equilibrium density matrix and $\chi_O$ is bond dimension of the non-evolved MPO), making such calculations prohibitively expensive in practice. In any case, the time-evolution of the operators $\hat{E}(x_0)$ lead to rather high bond-dimensions near $x = x_0$ within short times. While the associated truncations do not pose much of a problem for calculation of the linear Drude weight $\mathcal{D}^{(1)}_{\hat{j}_0}$ (see e.g. Refs. 19–22), we have found this to be inadequate even for calculations of $\mathcal{D}^{(2)}_{\hat{j}_0}$, where we were unable to obtain accurate values of $\chi^{(2)}_{\hat{j}_0}$ for large values of $(t_1 - t_0, t_2 - t_1)$ (the numerical experiments involving $\mathcal{D}^{(2)}_{\hat{j}_0}$ were performed on the easy-axis XXZ model with the Hamiltonian replaced by the energy current, which preserves integrability but breaks parity).

Therefore, in our numerical calculations we instead used the generalized Kohn formula[23,24] combined with exact diagonalization. The generalized Kohn formula relates the current Drude weights to the derivatives of the energy levels when a gauge flux $\varphi$ is threaded through a system with periodic boundary conditions. E.g. for $\mathcal{D}^{(3)}_{\hat{j}_0}$ it gives

$$\mathcal{D}^{(3)}_{\hat{j}_0} = \frac{1}{L}\sum_n p_n \frac{d^4\epsilon_n}{d\varphi^4} = \frac{1}{L}\sum_n p_n \frac{d^3\langle \hat{J}_0\rangle_n}{d\varphi^3}, \tag{F3}$$

where $L$ denotes the length of the system, $n$ runs over the eigenstates of $\hat{H}_0$, each of whom has energy $\epsilon_n$ and is occupied with probability $p_n$. In the second part $\hat{J}_0$ is the total charge current $\sum_j \hat{j}_0(j)$ and $\langle\cdot\rangle_n$ denotes the average over the $n$-th eigenstate. The figures reported in the main text are obtained by summing over all symmetry sectors (momentum and magnetization).

Note that a naive implementation of this formula based on finite differences would be problematic. For small enough $\varphi$ the numerical precision on the finite difference (which must the be divided by $\varphi^3$) would limit the accuracy of the results. On the other hand, at large enough $\varphi$, level crossings start to occur, thus compromising the results. Empirically, it seems that these two problems significantly compromise the results for all values of $\varphi$ starting at $L \gtrsim 15$.

There are two possible solutions to this problem. One is to use perturbation theory to express $\frac{d^4\epsilon_n}{d\varphi^4}$ based on matrix elements of $\hat{H}_0$ and $\hat{J}_0$ (see Eq. (31) of Ref. 24). Another alternative exploits the integrability of the model in question. In fact, we could choose a large $\varphi \simeq 10^{-2}$, and track levels through the various crossings based on their fidelity $\langle n(\varphi_0)|n(\varphi_1)\rangle$. Both approaches give consistent results for the cases we considered.

Finally, we point out that this approach is heavily limited by finite-size effects, specifically at small $\Delta$ or medium-high temperatures. Furthermore, in general, finite size effects are large enough to hinder a reliable extrapolation to the $L \to \infty$ limit. As an example, we report in Fig. S3 the finite size behaviour at smaller $\Delta$ as a function of temperature.

## Appendix G: Finite-field scaling in the XXX model

In this section we study the finite-field behaviour of the XXX chain and obtain the scaling form reported in Eq. (8) of the main text.

### 1. Infinite-temperature expansion of the TBA equations

Before proceeding to study the GHD equations, we summarise here the results from Ref. 25. Recall that the excitations of interest are quasiparticle "strings", which are bound states of $s$ elementary magnons. Throughout, we will assume that the length of the string $s$ is much smaller than $1/\mu = 1/(\beta h)$.

The occupation factors can be expanded in the $\beta \to \infty$ limit as

$$n_s = n_s^{(0)} + \beta n_s^{(1)} + O(\beta^2) \tag{G1}$$

where

$$n_0 \sim 1/s^2 \tag{G2}$$

asymptotically at large $s$. To obtain a response we need to have some rapidity-dependence in $n$, so we must consider at least $n^{(1)}$. Expanding the Yang-Yang equation, one can see that

$$n^{(1)} = -n^{(0)}\left(1 - n^{(0)}\right) e^{\mathrm{dr}}. \tag{G3}$$

This can be combined with the results

$$e^{\mathrm{dr}} = (k')^{\mathrm{dr}} = \frac{s+1}{2}\left(\frac{K_s}{s} - \frac{K_{s+1}}{s+2}\right) \tag{G4}$$

$$m^{\mathrm{dr}} \sim \frac{1}{3}\mu s^2, \tag{G5}$$

with

$$K_s(\theta) = \frac{s}{2\pi}\frac{1}{\theta^2 + \frac{s^2}{4}} \tag{G6}$$

Finally, recall that at infinite temperature $\partial_\theta$ and dressing commute.

### 2. Solution of the GHD equation

The equation we set out to solve is the inhomogeneous GHD equation (2), restricted to the case where $E$ and $n$ are uniform. Then, we have (see also Ref. 26)

$$\partial_t n_s - E a_s^{\mathrm{eff}} \partial_\theta n_s = 0. \tag{G7}$$

This equation can be solved through the method of characteristics (see e.g. Ref. 26). Introducing the vector potential

$$\varphi = -\int dt\, E(t), \tag{G8}$$

and using $\varphi$ to parametrize time we have

$$n_s(\theta, \varphi) = n_s(\theta_B(\varphi, \theta, s), 0), \tag{G9}$$

along the trajectory defined by

$$\frac{d\theta_B}{d\varphi} = -a_s^{\mathrm{eff}}, \tag{G10}$$

$$\theta_B(0, \theta, s) = \theta. \tag{G11}$$

Finally the $\varphi$-dependence of $\langle \hat{j}_0 \rangle$ is given by

$$\langle \hat{j}_0 \rangle = \sum_s j_s = \sum_s m_s^{\text{dr}} \int \frac{d\theta}{2\pi} n_s^{(1)}(\theta_B(\varphi,\theta))(e')_s^{\text{dr}}(\theta). \tag{G12}$$

We now proceed to show that at large $s$, but $\mu s \ll 1$, the expectation value of the current obeys the following asymptotic relation

$$j_s(\varphi) \sim \mu s^{-4} f(\mu s^3 \varphi), \tag{G13}$$

for some function $f$. We consider the set of equations (G10), (G11), and (G12), and perform the rescaling

$$\lambda = \theta s^{-1}, \tag{G14}$$
$$\lambda_B = \theta_B s^{-1}, \tag{G15}$$
$$\phi = \mu \varphi s^3. \tag{G16}$$

Plugging this into (G10), we have

$$\frac{d\lambda_B}{d\phi} = \frac{m_s^{\text{dr}}}{\mu s^2} \frac{1}{s^2 (k')_s^{\text{dr}} (s\lambda_B)}. \tag{G17}$$

It is now easy to check that both ratios on the RHS of this equation tend to a well defined limit for large $s$ (while holding $\lambda_B = \mathcal{O}(1)$ fixed). Thus we see that $\lambda_B(\phi)$ will be $s$-independent.

We can now apply the same rescaling to Eq. (G12), where we get

$$j_s(\phi) = -\underbrace{\beta n_s^{(0)} \left(1 - n_s^{(0)}\right) m_s^{\text{dr}}}_{=\mu \mathcal{O}(1)} \int \frac{s d\lambda}{2\pi} \underbrace{e_s^{\text{dr}}(s, s\lambda_B(\phi))}_{s^{-2} \mathcal{O}(1)} \underbrace{(e')^{\text{dr}}(s, s\lambda)}_{s^{-3} \mathcal{O}(1)}, \tag{G18}$$

Regrouping all the $s$-dependencies and denoting the remaining $\mathcal{O}(1)$ function by $G$, we have

$$j_s(\varphi) \sim \mu s^{-4} \int d\lambda G(\lambda, \varphi) \sim \mu s^{-4} f\left(\mu s^3 \varphi\right), \tag{G19}$$

for some function $f$. Finally, note that $f$ must be odd and periodic with some period $\zeta$. The first observation follows from parity, under which both $\phi$ and $\hat{j}$ are odd. The periodicity, instead follows from the following result. For any $s$ there is a value of $\phi$ for which $\theta_B(\phi,\theta,s) = \theta \,\forall \theta$, thus $j_s(\phi)$ must be periodic, and consequently so must $f$.

To show this last statement, we note that Eq. (G10) is separable. Denoting the primitive of $(k')_s^{\text{dr}}$ by $I_s(\theta)$, we have

$$I_s(\theta_B(\varphi,\theta,s)) = -\varphi m_s^{\text{dr}} + I_s(\theta), \tag{G20}$$

until the time $\theta_B$ reaches $+\infty$. At that point, recalling the periodic boundary condition in rapidity space (see e.g. Ref. 26), we can imagine that $\theta_B$ is transported to $-\infty$. Thereafter the evolution is described by

$$I_s(\theta_B(\varphi,\theta,s)) = -\varphi m_s^{\text{dr}} + I_s(\theta) + (I_s(+\infty) - I_s(-\infty)). \tag{G21}$$

From this last equation we can immediately notice that for

$$\varphi = \frac{I_s(+\infty) - I_s(-\infty)}{m_s^{\text{dr}}}, \tag{G22}$$

$\theta_B(\phi,\theta,s) = \theta \,\forall \theta$, as claimed.

---



\end{document}